\documentclass[journal,10pt]{IEEEtran}
\usepackage[utf8]{inputenc}
\usepackage{array}
\usepackage[nolist]{acronym}
\usepackage{amsmath}
\usepackage{amssymb}
\usepackage{cases,cite}
\usepackage{mathtools}
\usepackage{psfrag}
\usepackage{pstool}
\usepackage{pgfplots}
\usepackage{xcolor}
\usepackage{tikz}
\usepackage{gensymb} 
\usepackage{siunitx}
\usepackage{pifont}

\usepackage{balance}
\usepackage{graphicx}
\usepackage{blindtext, graphicx}   
\usepackage{booktabs,tabularx}
\usepackage{colortbl}
\definecolor{LightCyan}{rgb}{0.88,1,1}
\newcolumntype{L}{>{\centering\arraybackslash}m{3cm}}
\usepackage[urlcolor=blue]{hyperref} 

\usepackage[font=small,skip=0pt]{caption}
\usepackage{enumerate}
\usepackage[list=true]{subcaption}
\newcommand{\figref}[1]{\figurename~\ref{#1}}
\usepackage[ruled,vlined]{algorithm2e}
\usepackage{algorithmic}

\newcommand{\row}{g}
\newcommand{\col}{v}
\newcommand{\wpq}{{\mathbf{W}}_{\row \col}}

\newcommand{\eirpc}{{\mathrm{EIRP}}_{\mathrm{c}}}
\newcommand{\eirpd}{{\mathrm{EIRP}}_{\mathrm{d}}}

\newcommand{\sllmin}{{\mathrm{SLL}}_{\mathrm{min}}}

\newcommand{\slla}{{\mathrm{SLL}}^{(\mathrm{a})}}
\newcommand{\sllac}{{\mathrm{SLL}}_{\mathrm{c}}^{(\mathrm{a})}}
\newcommand{\sllad}{{\mathrm{SLL}}_{\mathrm{o}}^{(\mathrm{a})}}

\newcommand{\slle}{{\mathrm{SLL}}^{(\mathrm{e})}}
\newcommand{\sllec}{{\mathrm{SLL}}_{\mathrm{c}}^{(\mathrm{e})}}
\newcommand{\slled}{{\mathrm{SLL}}_{\mathrm{o}}^{(\mathrm{e})}}

\newcommand{\thetan}{\theta_\mathrm{Null}}
\newcommand{\phin}{\phi_\mathrm{Null}}

\newcommand{\ppe}{\mathrm{PPE}}

\newcommand{\thetat}{{\theta_{3\text{dB}}}}
\newcommand{\thetae}{\theta_{3\text{dB}}^{(\mathrm{e})}}
\newcommand{\thetaa}{\theta_{3\text{dB}}^{(\mathrm{a})}}

\SetCommentSty{mycommfont}

\SetKwInput{KwInput}{Input}                
\SetKwInput{KwOutput}{Output}              

\usetikzlibrary{shapes.geometric, arrows}
\tikzstyle{startstop} = [rectangle, rounded corners, 
minimum width=1cm, 
minimum height=2cm,
text centered, 
text width=1cm,
draw=black, 
fill=red!30]

\tikzstyle{io} = [trapezium, 
trapezium stretches=true, 
trapezium left angle=70, 
trapezium right angle=110, 
minimum width=3cm, 
minimum height=1cm, text centered, 
draw=black, fill=blue!30]

\tikzstyle{process} = [rectangle, 
minimum width=1cm, 
minimum height=2cm, 
text centered, 
text width=1cm, 
draw=black, 
fill=orange!30]

\tikzstyle{decision} = [diamond, 
minimum width=3cm, 
minimum height=1cm, 
text centered, 
draw=black, 
fill=green!30]
\tikzstyle{arrow} = [thick,->,>=stealth]

\begin{document}
\bstctlcite{IEEEexample:BSTcontrol}
\title{A Deep-NN Beamforming Approach for \\ Dual Function Radar-Communication THz UAV}  

\author{Gianluca Fontanesi, \IEEEmembership{Member, IEEE} Anna Guerra, \IEEEmembership{Member, IEEE}, Francesco Guidi, \IEEEmembership{Member, IEEE}, \\ Juan A. V{\'a}squez-Peralvo, \IEEEmembership{Member, IEEE}, Nir Shlezinger, \IEEEmembership{Member, IEEE}, Alberto Zanella, \IEEEmembership{Senior Member, IEEE}, Eva Lagunas, \IEEEmembership{Senior Member, IEEE}, Symeon Chatzinotas, \IEEEmembership{Fellow, IEEE}, Davide Dardari, \IEEEmembership{Senior Member, IEEE}, and Petar M. Djuri\'c, \IEEEmembership{Life Fellow, IEEE}
\thanks{G. Fontanesi was with the University of Luxembourg when the work was started and it is currently with Nokia Bell Labs, Germany \{gianluca.fontanesi@nokia.com\}. \\
J. V{\'a}squez-Peralvo, E. Lagunas and S. Chatzinotas are with the University of Luxembourg \{juan.vasquez, eva.lagunas,symeon.chatzinotas@uni.lu\}.\\
A. Guerra, F. Guidi and A. Zanella are with the National Research Council of Italy (\{anna.guerra, francesco.guidi, alberto.zanella\}@cnr.it)\\
N. Shlezinger is with Ben Gurion University, Israel. (nirshl@bgu.ac.il). \\
D. Dardari is with WiLab-CNIT/DEI, University of Bologna, Italy. (davide.dardari@unibo.it).\\
P. M. Djuri\'c is with the Stony Brook University, New York, USA. (petar.djuric@stonybrook.edu).
}
}
\markboth{submitted to IEEE Transactions on Vehicular Technology}
{}

\maketitle

\begin{abstract}
In this paper, we consider a scenario with one \ac{UAV} equipped with a \ac{ULA}, which sends combined information and sensing signals to communicate with multiple \acp{GBS} and, at the same time, senses potential targets placed within an interested area on the ground. We aim to jointly design the transmit beamforming with the \acp{GBS} association to optimize communication performance while ensuring high sensing accuracy. 
We propose a predictive beamforming framework based on a dual \ac{DNN} solution to solve the formulated nonconvex optimization problem. A first \ac{DNN} is trained to produce the required beamforming matrix for any point of the \ac{UAV} flying area in a reduced time compared to state-of-the-art beamforming optimizers. A second \ac{DNN} is trained to learn the optimal mapping from the input features, power, and \ac{EIRP} constraints to the \acp{GBS} association decision. Finally, we provide an extensive simulation analysis to corroborate the proposed approach and show the benefits of \ac{EIRP}, \ac{SINR} performance and computational speed.
\end{abstract}

\IEEEpeerreviewmaketitle

\newcommand{\targ}{0}
\newcommand{\dest}{i}
\newcommand{\bsk}{k}
\newcommand{\bskn}{\hat{k}_n}
\newcommand{\instant}{n}
\newcommand{\ant}{m}
\newcommand{\dt}{\delta_t}
\newcommand{\uk}{\mathbf{u}_{\bsk}}
\newcommand{\ui}{\mathbf{u}_{\dest}}
\newcommand{\ei}{\mathbf{e}_{\targ}}
\newcommand{\xii}{x_{\dest}}
\newcommand{\yi}{y_{\dest}}
\newcommand{\zi}{z_{\dest}}
\newcommand{\hui}{\hat{\boldsymbol{u}}_{\dest}}
\newcommand{\ukn}{\mathbf{u}_{\bskn}}
\newcommand{\xk}{x_{\bsk}}
\newcommand{\yk}{y_{\bsk}}
\newcommand{\zk}{z_{\bsk}}
\newcommand{\hk}{h_{\bsk}}
\newcommand{\dk}{d_{\bsk}}
\newcommand{\di}{d_{\dest}}
\newcommand{\thetak}{\theta_{\bsk}(\instant)}
\newcommand{\thetai}{\theta_{\dest}}
\newcommand{\phik}{\phi_{\bsk}(\instant)}
\newcommand{\phii}{\phi_{\dest}(\instant)}
\newcommand{\qn}{\mathbf{q}(\instant)}
\newcommand{\xn}{x(\instant)}
\newcommand{\yn}{y(\instant)}
\newcommand{\zn}{z(\instant)}
\newcommand{\Vmax}{V_{\text{max}}}
\newcommand{\poss}{\mathbf{u}_{\targ}}
\newcommand{\xs}{x_{\targ}}
\newcommand{\ys}{y_{\targ}}
\newcommand{\zs}{z_{\targ}}
\newcommand{\posm}{\mathbf{p}_{\ant}}
\newcommand{\pmo}{\mathbf{p}_{\ant}^{(0)}}
\newcommand{\xm}{x_{\ant}}
\newcommand{\ym}{y_{\ant}}
\newcommand{\zm}{z_{\ant}}
\newcommand{\dant}{d_{\text{ant}}}
\newcommand{\Rot}{\mathbf{R}}
\newcommand{\Rotx}{\mathbf{R}_x}
\newcommand{\Roty}{\mathbf{R}_y}
\newcommand{\Rotz}{\mathbf{R}_z}
\newcommand{\fc}{f_\mathrm{c}}
\newcommand{\Thetamin}{\Theta_{\text{min}}}
\newcommand{\Thetamax}{\Theta_{\text{max}}}
\acresetall
\section{Introduction}
\acresetall
\IEEEPARstart{W}{ireless} communications and radio sensing are evolving towards the same technological solutions involving high frequencies, large antenna arrays, and miniaturized devices \cite{SaaBenChe:J20,wei2022toward}. Thereby, integrating sensing capabilities in wireless infrastructures offers new exciting opportunities for the next \ac{6G} cellular systems and beyond \cite{WilBraVis:J21,liu2022integrated}.
However, sensing and communications have different roles: sensing collects and extracts information from noisy data, whereas communications are devoted to transmitting information through ad-hoc signaling schemes and recovering it from a noisy environment.  The \ac{ISAC} paradigm integrates both functionalities (e.g., by using the same hardware) to find a trade-off between competing needs and mutual performance gains \cite{chiriyath2017radar_commun, liu2020joint}. 

Several works have recently proposed different waveform designs to find the best sensing and communication trade-offs.
In \cite{chalise2017performance}, the authors consider the trade-off between the detection probability and achievable rate for a joint communication and passive radar system. Then, \cite{chalise2018performance} accounts for the trade-off between estimation and communication and the design of a waveform favorable for target estimation and information delivery. Finally, in \cite{liu2021cramer}, the authors investigate the trade-off arising in \ac{ISAC} systems due to the different treatment of spatial \acp{DOF}. 

Waveform design plays a key role in attaining  integration gain and can be conceived either in a non-overlapping resource allocation scheme or in a fully unified framework \cite{liu2022integrated}. In the first case, sensing and communications are split over orthogonal (non-overlapping) wireless resources \cite{liu2022integrated,ma2020joint}. Differently, fully unified waveform design can follow two approaches: (1) a sensing-centric scheme when a typical sensing waveform (e.g., chirp signals) incorporates a communication functionality, e.g., \cite{huang2020majorcom,ma2021frac}; (2) a communication-centric scheme when a communication waveform (e.g., OFDM) is also used for sensing \cite{chen2021code}, and a joint design approach based on optimization \cite{johnston2022mimo}. 

Such a joint design is particularly appealing for networks of low-complexity devices, such as \acp{UAV}, as it allows the design of highly flexible and efficient systems overcoming the size, weight and endurance constraints of autonomous agents \cite{zhang2021overview, wu2021comprehensive}. 
Indeed, towards the realization of \ac{6G}, \acp{UAV} have attracted significant interest thanks to their low cost and their flexibility, which let them be a suitable solution for both communication and sensing \cite{guerra2020dynamic,MerGuv:C15,FonEtAl:J23,guerra2020dynamic-2}. 
\acp{UAV} are emerging sensing technologies that, thanks to their flexibility and their possibility of keeping a privileged \ac{LoS} point of view, are often used for localization and sensing in time-critical applications  \cite{wang2021multi,GueEtAl:J22, zhang2020self,wang2019autonomous}.
The advantages of using \acp{UAV} with \ac{ISAC} are many, but at the extreme, one can find two significant aspects. On the one hand, the \acp{UAV}' 3D mobility allows \ac{DFRC} tasks to be performed in an optimized manner: an optimized \acp{UAV} trajectory can further increase the performance of the \ac{ISAC} system. On the other hand, \ac{ISAC} is an integrated solution that supports easy and low-complexity hardware deployment of onboard battery constraint agents that requires determining the best beamforming and waveform design.
In this perspective, high-frequency technology, such as \ac{THz} and mm-Wave, has emerged as a promising solution for sensing \cite{lotti2022radio}, for \ac{UAV} integration thanks to their ease of miniaturization \cite{GueEtAl:C21}, and for all the solutions that entail the convergence of communication, localization and sensing capabilities \cite{sarieddeen2020next}. 

\subsection{Related Works and Contributions}
The \ac{ISAC} theoretical framework has been recently applied to \acp{UAV} \cite{meng:Traject_Beamf_CommLetter2022, meng2022throughput, zhang2021uav, jing2022isac, wang2022reinforcement, hu2022trajectory, zhang2018joint, lyu:jointTrajeBeam_ICC2022,10168298}, which can be classified according to the UAV's trajectory constraints and goals. 
More specifically, \cite{jing2022isac, hu2022trajectory, zhang2018joint} consider the energy consumption of the \ac{UAV} during the trajectory while achieving the sensing performance gains.
In \cite{jing2022isac}, a rotatory-wing \ac{UAV} transmits the \ac{ISAC} signal during its flight to simultaneously provide downlink communication service to a ground user and sense a target. 
The trajectory design problem aims to determine the subsequent \ac{UAV} waypoints, hover points, and flight speeds to maximize the average communication rate while minimizing the \ac{CRLB} of the target location estimate. In \cite{hu2022trajectory}, a \ac{GBS} is deployed to deliver downlink wireless services to cellular users. A cellular-connected \ac{UAV} equipped with a side-looking \ac{SAR} flies and collects the echoes of communication signals originating from the \ac{GBS} to sense objects and gain situational awareness. 
The \ac{UAV} minimizes the overall propulsion energy consumption during the time horizon while maintaining acceptable sensing resolution by reusing cellular communication signals.
In \cite{zhang2018joint}, the considered \ac{UAV} is required to execute multiple sensing tasks in sequence within the cell coverage.
Nevertheless, in \cite{zhang2018joint} it has not been considered that, due to its high altitude, the \ac{UAV} might associate with several candidates \acp{GBS} at different distances.
Thus, the \ac{UAV}-\ac{GBS} association should be carefully considered when designing the \ac{UAV} trajectory.

Other works as \cite{lyu:jointTrajectBeam_Journal2022,lyu:jointTrajeBeam_ICC2022, meng:Traject_Beamf_CommLetter2022, hua2023optimal_TVT, ren:optimalSecrecyBeamform_ICC2022, Guerra2023_TAES_RL_UAV_networks} focus on the \ac{UAV} beamforming problem considering antenna arrays deployed on the \ac{UAV}.
The authors in \cite{lyu:jointTrajectBeam_Journal2022,lyu:jointTrajeBeam_ICC2022} consider a \ac{UAV} equipped with a \ac{ULA} to serve ground users and, simultaneously, to perform radar sensing towards potential ground targets.
The objective is to maximize the average weighted sum-rate throughput by jointly optimizing the \ac{UAV} trajectory, as well as the transmit information and sensing beamforming subject to the sensing requirements and transmit power constraints over different time slots.

In \cite{meng:Traject_Beamf_CommLetter2022}, the authors consider a \ac{UAV}-\ac{ISAC} system integrated with a \ac{ULA}, to maximize the achievable rate, subject to the beam-pattern gain constraint and the maximum transmit power constraint.
Works \cite{hua2023optimal_TVT, ren:optimalSecrecyBeamform_ICC2022} minimize a beampattern matching error by jointly optimizing the sensing and communication beamforming design subject to the transmit power constraints and secrecy rate \cite{ren:optimalSecrecyBeamform_ICC2022} or transmit power and rate constraints \cite{hua2023optimal_TVT}.

The antenna pattern in these papers is the result of optimization solutions. However, simultaneously satisfying SLL, beamwidth, nulling and EIRP constraints is difficult in these approaches. This becomes even more complicated when a realistic high gain sub-array is considered during the beam steering process.

In addition, the rotation of the antenna radiation pattern suffered by small/medium \acp{UAV} due to environmental factors has not yet been considered when considering the effectiveness of the proposed beamforming patterns.

These joint challenges introduce a high level of complexity. First, a realistic pattern synthesizer is 
complicated for the formed 3D beam based on a ULA in real-time ISAC application scenarios, especially in \ac{UAV}-enabled ISAC networks where wireless links have diverse and varying elevation and azimuth angles.
The complexity is even increased because the azimuth and elevation angles are not influenced for each trajectory point but are made from the rotation of the \ac{UAV} and the applied \ac{GBS} association policy.

This work aims to propose a \ac{DNN} solution that, together with the conventional challenges of \ac{ISAC}, takes into account the \ac{UAV}-\ac{GBS} association problem described above, a realistic antenna beam pattern synthesizer and the possible rotation of the radiation pattern.
To more efficiently exploit the benefits of beamforming in \ac{UAV}-enabled ISAC networks, a more flexible antenna array, e.g., an \ac{ULA}, is considered to form dedicated beams.

Thus, the main contributions of this manuscript can be summarized as follows:
\begin{itemize}
    \item We consider a \ac{THz} cellular-connected \ac{UAV} system, required to communicate with surrounding \acp{GBS} and perform target sensing operation. Unlike previous works, we consider a realistic on-board \ac{UAV} antenna design and beamforming, controlling the beam's position, the side lobe levels, and the nulling. In addition, we consider the rotation of the \ac{ULA} antenna during the \ac{UAV} trajectory.
    \item We formulate a novel \ac{UAV} trajectory problem based on joint beamforming design and GBS association policy.
    The complexity of the proposed real-time antenna synthesizer, coupled with the nonconvex association constraints and \ac{UAV} antenna rotation, makes the formulated problem nonconvex and high-dimensional. Consequently, conventional antenna synthesizers and mathematical approaches are inapplicable.  
    \item We propose a double \ac{DNN} based approach to approximate the non-linear mapping from the \ac{UAV} position to the optimal beamforming weights selection and the optimal \ac{GBS} selection.
    The labeled radio data for the first \ac{DNN} is generated using an antenna pattern optimizer, which can synthesize a radiation pattern considering the beamwidth, pointing direction, and nulling as input data. 
    \item With the advantage of reducing the time to generate the required beam, our \ac{DNN} solution learns the directional beamforming weights for each \ac{UAV}'s position in the flight area, including any effect on elevation and azimuth angle due to \ac{UAV} rotation. In addition, a second \ac{DNN} reduces the signaling overhead for the \ac{GBS}-\ac{UAV} association, indicating the best \ac{GBS} association in terms of rate and minimum EIRP.
    \item We provide an extensive simulation analysis to corroborate the proposed framework and show its effectiveness in accuracy and speed compared to the antenna optimizer.
\end{itemize}

The rest of the paper is organized as follows: Sec.~\ref{SystemModel} introduces the system model of the \ac{DFRC} \ac{UAV} system. In Sec.~\ref{ProblemFormulation}, we formulate the connectivity-constrained beam-pattern gain minimization problem, while Sec.~\ref{sec:proposed} details the proposed \ac{NN}-aided approach. Numerical results are reported in Sec.~\ref{sec:sims}, while Sec.~\ref{sec:conclusions} provides concluding remarks. 

\begin{figure}[t!]
\centering
\input{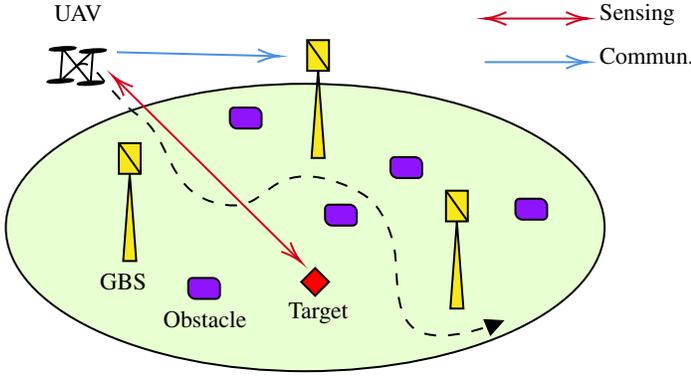}
\vspace{0.1cm}
    \caption{Considered scenario, where the \ac{DFRC} \ac{UAV} needs to autonomously navigate the environment while guaranteeing reliable communication and sensing performance.}  \label{fig:scenario}
\end{figure}
A complete list of recent works on ISAC systems applied to UAV technology is reported in Table \ref{tab:Related_Works}. 

\begin{table*}[t]
\centering
\begin{tabularx}{\linewidth}{|c|c|c|c|X|X|}
\hline
\rowcolor{LightCyan}
\hline
Ref.  &  ISAC &   \ac{UAV} Role   &  Optimization Goal   & Constraints &  Opt. Technique          \\
\hline
\cite{jing2022isac} & No&  \ac{UAV}-BS  & Communication Rate, CRB  & Energy, \ac{UAV} mobility & Iterative Algorithm \\ 
\hline
\cite{hu2022trajectory} & Yes  & \ac{UAV}-UE   &  Tx Power, Range Resolution  & Overall propulsion energy consumption & BCD  \\
\hline
\cite{lyu:jointTrajeBeam_ICC2022} & Yes    & \ac{UAV}-BS  & Comm. Rate & Sensing Beam-Pattern Gain, \ac{UAV} TX Power & SCA, SDR \\
\hline
\cite{zhang2018joint} & No &  \ac{UAV}-UE & Tx Power, \ac{UAV} speed & Velocity, acceleration, \ac{UAV} Coverage, Sensing Rate & Problem decomposition\\
\hline
\cite{lyu:jointTrajectBeam_Journal2022} & Yes    & \ac{UAV}-BS  & Comm. Rate & Sensing Beam-Pattern Gain, \ac{UAV} TX Power & CVX \\
\hline
\cite{meng:Traject_Beamf_CommLetter2022} & Yes   & \ac{UAV}-BS & Communication Rate &  Beam-Pattern Gain, max Transmit Power & SDR, eigenvalue decomposition\\
\hline
\cite{hua2023optimal_TVT} & Yes & / & Beampattern error & Transmit Power, Rate & SDR \\
\hline
\cite{ren:optimalSecrecyBeamform_ICC2022} & Yes & / & Beampattern error &  Transmit Power, Secrecy Rate & SDR \\
\hline
This work & Yes & \ac{UAV}-BS/UE & Beampattern error  &  Transmit Power, GBS Connectivity & DNN \\
\hline
\end{tabularx}
\caption {Related works on \ac{UAV}-ISAC systems.}
\label{tab:Related_Works}
\end{table*}

\section{System Model}\label{SystemModel}

We consider a cellular system, schematically depicted in \figref{fig:scenario}, populated by a terrestrial network of \acp{GBS} and a \ac{UAV}. 
More specifically, the \ac{UAV} is required to act as a mobile \ac{DFRC} system that simultaneously probes its surrounding environment through a monostatic radar while communicating with its associated \ac{GBS}.
We aim to facilitate the operation of such networks by adequately selecting the \ac{GBS} and the choice of the suitable orientation and beam-pattern configuration for the \ac{UAV} through an ad-hoc beam-pattern optimization technique to satisfy communication and sensing requirements. 

Next, we first discuss the \ac{GBS} association rule. After providing  a brief overview of the considered geometry, we present the transmitted signal model and describe how we consider the signaling for sensing and communication, the ground-to-air channel, and finally, the received signal model. 

\subsection{GBS Association Policies}\label{BS_Association}
Selecting the best \ac{GBS} to associate with is not trivial due to blockages, \ac{NLoS} links, or communication/sensing pointing angles that make the radiation pattern optimization unsuitable.

In particular, one can consider one of the following association approaches, defined also as policy in the results:
\begin{enumerate}
    \item A possible association rule is represented by choosing the closest \ac{GBS}. However, if the nearest-neighbor association is adopted, performance might degrade due to blockages. In addition, the communication and sensing angle might be very different.
    \item A second possibility is to select the \ac{GBS} with an azimuth angle close to the \ac{UAV}-target azimuth angle to maximize the sensing performance.
    \item A third possibility is to select the \ac{GBS} experiencing the highest \ac{SINR}. This requires a complete scan of the environment to receive the \ac{GBS} pilot information and, thus, higher overhead and latency. In addition, it might happen that the \ac{GBS} with the highest \ac{SINR} is not the one that allows optimizing the beam pattern for joint sensing and communication purposes.
\end{enumerate}
As noted above, each approach has its challenges and limitations. 
Thus, we will investigate the performance of the three approaches for the system's joint sensing and communication performance.

\subsection{System Geometry}

The \ac{UAV} mission period, namely $\mathcal{T} = [0,T]$ is discretized into $N$ time slots, each with duration $\dt = T/N$. 
Here, $\dt$ is chosen to be sufficiently small so that the \ac{UAV} location can be assumed to be approximately unchanged within each slot to facilitate the trajectory and beamforming design.
In addition, we consider that the \ac{UAV} aims to sense $U=1$ target of interest at a known location $\poss = (\xs, \ys, \zs)$. At any time slot during the \ac{UAV} trajectory, the \ac{UAV} can associate only with one of the \ac{GBS}.

\begin{figure}[!t]
\psfrag{X}[c][c][0.8]{$X$}
\psfrag{Y}[c][c][0.8]{$Y$}
\psfrag{Z}[c][c][0.8]{$Z$}
\psfrag{uk}[c][c][0.8]{$\uk$}
\psfrag{i}[c][c][0.8]{$ \vec{i}$}
\psfrag{tm}[c][c][0.8]{\quad$\tau_{\ant}$}
\psfrag{t}[c][c][0.8]{$\tau$}
\psfrag{pm}[c][c][0.8]{$\posm^{(0)}$}
\psfrag{qn}[rc][rc][0.8]{$\qn$}
\psfrag{tk}[c][c][0.8]{$\thetak$}
\psfrag{pk}[lc][lc][0.8]{$\phik$}
\psfrag{a}[lc][lc][0.8]{$\alpha$}
\psfrag{b}[lc][lc][0.8]{$\beta$}
\psfrag{g}[lc][lc][0.8]{$\gamma$}
\centering
\includegraphics[width=0.95\linewidth,draft=false]{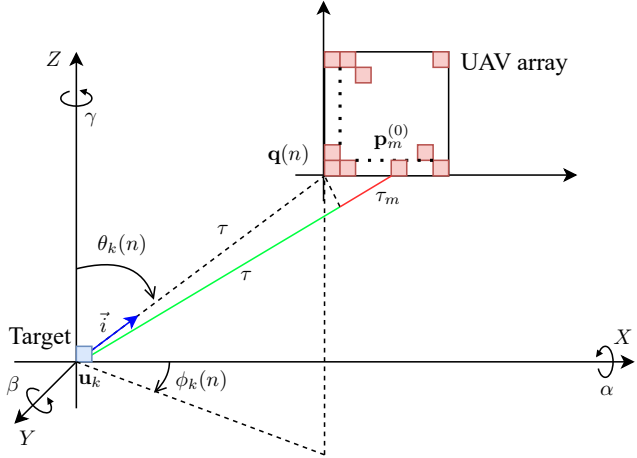}
\caption{System scenario and geometry. 
}\label{fig:geometry}
\end{figure}

 We consider a 3D Cartesian system, represented in Fig. \ref{fig:geometry}, where 
\begin{itemize}
    \item The location of the \acp{GBS} are fixed at $\uk=(\xk, \yk, \zk), \quad \bsk = {1,...,K}$ with $K$ being the number of deployed \acp{GBS}, $(\xk, \yk)$ denoting the location of the $\bsk$th \ac{GBS} and $\zk  = h_{\text{BS}},\, \forall k$;
    \item The time-varying location of the \ac{UAV} at time slot $n \in \mathcal{N}$ is $\qn = (\xn,\, \yn, \zn)$. Then, consider the set of \ac{UAV} trajectories as 
\begin{align}
\mathcal{Q} = \{\mathbf{q}_1,\,\ldots,\,\mathbf{q}_{\ell},\,\ldots,\,\mathbf{q}_{N_\mathrm{T}}\},
\end{align}
where each trajectory can be expressed as a sequence of UAV positions
\begin{align}
\mathbf{q}_{\ell} =\{\mathbf{q}_{\ell}(1), \ldots, \mathbf{q}_{\ell}(n), \ldots, \mathbf{q}_{\ell}(N)\},
\end{align}
and $N_\mathrm{T}$ is the total number of flights performed by the \ac{UAV}.
We consider the initial and final positions being fixed, that is, $\mathbf{q}_\ell(1)=\mathbf{q}_{\mathrm{I}}$, $ \mathbf{q}_\ell(N)=\mathbf{q}_{\mathrm{F}}$, $\forall\,\ell$.
    \item The \ac{UAV} speed is fixed and upper limited by $\lVert \mathbf{q}(\instant+1)-\qn\rVert \leq \dt\, \Vmax$ with $\Vmax$ being the maximum speed. 
\end{itemize}

The \ac{UAV} is equipped with an array of $M$ antennas centered in $\qn$ whose coordinates are
\begin{align}\label{eq:antenna}
    &\posm= \qn +(\xm,\, \ym,\, \zm)=\qn+\Rot(\alpha, \beta, \gamma)\, \pmo,
\end{align}
with $m \in \left\{0,1,\ldots, M-1\right\}$, and where 
\begin{align}
 \Rot(\alpha, \beta, \gamma) = \Rotx(\alpha)\Roty(\beta) \Rotz(\gamma)   
\end{align}
is a generic 3D rotation matrix with $(\alpha, \beta, \gamma)$ being the rotational angles around the $x$, $y$, and $z$ axis respectively, and $\pmo$ is the antenna array position before the rotation
 { The definition of the rotation matrices $\Rotx(\alpha)\Roty(\beta) \Rotz(\gamma)$ is defined in \cite{guerra:SingleAnchorTWCOM2018},\cite[3.42]{lavalle2006planning} and it is reported in \eqref{eq:rot}. Note that the orientation is affected by the trajectory only and not from speed variation, that we leave for future work. Consequently, it is straightforward to find the relationships in \eqref{eq:coordinates} between the coordinates of the antenna elements and the rotational angles.
\begin{figure*}
\begin{align}\label{eq:rot}
    &\Rot(\alpha, \beta, \gamma) = \left[
    \begin{array}{ccc}
    \cos(\alpha)\cos(\beta) & \cos(\alpha)\sin(\beta) \sin(\gamma) - \sin(\alpha) \cos(\gamma) & \cos(\alpha)\sin(\beta)\cos(\gamma) + \sin(\alpha) \sin(\gamma) \\
    \sin(\alpha)\cos(\beta) &  \sin(\alpha)\sin(\beta) \sin(\gamma) +\cos(\alpha) \cos(\gamma) & \sin(\alpha)\sin(\beta)\cos(\gamma) - \cos(\alpha) \sin(\gamma)  \\
    -\sin(\beta) & \cos(\beta) \sin(\gamma) & \cos(\beta) \cos(\gamma)
\end{array}
\right],
\end{align}
\end{figure*}
\begin{figure*}
\begin{align}\label{eq:coordinates}
    &x_m=x_m^{(0)} \cos(\alpha)\cos(\beta) + y_m^{(0)} \left[\cos(\alpha) \sin(\beta) \sin(\gamma) - \sin(\alpha) \cos(\gamma) \right] + z_m^{(0)} [\cos(\alpha) \sin(\beta) \cos(\gamma) + \sin(\alpha) \sin(\gamma)] , \\
    &y_m=x_m^{(0)} \sin(\alpha)\cos(\beta)+ y_m^{(0)} \left[\sin(\alpha) \sin(\beta) \sin(\gamma) + \cos(\alpha) \cos(\gamma) \right] + z_m^{(0)} \left[\sin(\alpha) \sin(\beta) \cos(\gamma) - \cos(\alpha) \sin(\gamma) \right] , \\
    &z_m= -x_m^{(0)} \sin(\beta) + y_m^{(0)} \cos(\beta) \sin(\gamma) + z_m^{(0)} \cos(\beta) \cos(\gamma) .
\end{align}
\end{figure*}
}

Note that eq.~\eqref{eq:antenna} can specialize in different array geometries. In this work, we consider the \ac{UAV} equipped with a squared \ac{UPA} with $M$ antennas, initially distributed along the $XZ$-plane, such that it holds
\begin{align}\label{eq:antennaULAY}
    &\posm^{(0)}= (\xm,\, \ym,\, \zm)=\left( \ant_x\, \frac{\lambda}{2}, 0,\,\, \ant_z\, \frac{\lambda}{2}\right),
\end{align}
where 
\begin{align}
\ant_x &= \left(1+\left\lfloor\frac{m}{\sqrt{M}}\right\rfloor\right) \frac{\lambda}{2}, \nonumber \\ 
\ant_z &= \left[ 1+ (m \! \mod \sqrt{M} )\right]\frac{\lambda}{2} \nonumber 
\end{align}
and we have assumed that the inter-antenna spacing is $\dant=\frac{\lambda}{2}$ with $\lambda$ being the carrier wavelength. In the sequel, we consider an optimization problem to control the array orientation using the rotational operation defined by \eqref{eq:antenna}.

\subsection{Transmitted Signal Model}

\begin{figure}
\centering
  
\tikzset {_07y1vsxxi/.code = {\pgfsetadditionalshadetransform{ \pgftransformshift{\pgfpoint{0 bp } { 0 bp }  }  \pgftransformrotate{0 }  \pgftransformscale{2 }  }}}
\pgfdeclarehorizontalshading{_xo1jyet8b}{150bp}{rgb(0bp)=(0.96,0.88,0.99);
rgb(37.589285714285715bp)=(0.96,0.88,0.99);
rgb(62.5bp)=(0.95,1,0.88);
rgb(100bp)=(0.95,1,0.88)}
\tikzset{every picture/.style={line width=0.75pt}} 
\tikzset{every picture/.style={font=\small}}

\begin{tikzpicture}[x=0.75pt,y=0.75pt,yscale=-0.7,xscale=0.7]

\draw  [fill={rgb, 255:red, 230; green, 255; blue, 204 }  ,fill opacity=1 ] (10.5,17.71) .. controls (10.5,12.35) and (14.85,8) .. (20.21,8) -- (118.79,8) .. controls (124.15,8) and (128.5,12.35) .. (128.5,17.71) -- (128.5,46.84) .. controls (128.5,52.21) and (124.15,56.55) .. (118.79,56.55) -- (20.21,56.55) .. controls (14.85,56.55) and (10.5,52.21) .. (10.5,46.84) -- cycle ;

\draw  [fill={rgb, 255:red, 230; green, 255; blue, 204 }  ,fill opacity=1 ] (170.5,17.71) .. controls (170.5,12.35) and (174.85,8) .. (180.21,8) -- (278.79,8) .. controls (284.15,8) and (288.5,12.35) .. (288.5,17.71) -- (288.5,46.84) .. controls (288.5,52.21) and (284.15,56.55) .. (278.79,56.55) -- (180.21,56.55) .. controls (174.85,56.55) and (170.5,52.21) .. (170.5,46.84) -- cycle ;

\draw  [fill={rgb, 255:red, 243; green, 227; blue, 255 }  ,fill opacity=1 ] (10.5,94.16) .. controls (10.5,88.79) and (14.85,84.45) .. (20.21,84.45) -- (118.79,84.45) .. controls (124.15,84.45) and (128.5,88.79) .. (128.5,94.16) -- (128.5,123.29) .. controls (128.5,128.65) and (124.15,133) .. (118.79,133) -- (20.21,133) .. controls (14.85,133) and (10.5,128.65) .. (10.5,123.29) -- cycle ;

\draw  [fill={rgb, 255:red, 243; green, 227; blue, 255 }  ,fill opacity=1 ] (170.5,94.16) .. controls (170.5,88.79) and (174.85,84.45) .. (180.21,84.45) -- (278.79,84.45) .. controls (284.15,84.45) and (288.5,88.79) .. (288.5,94.16) -- (288.5,123.29) .. controls (288.5,128.65) and (284.15,133) .. (278.79,133) -- (180.21,133) .. controls (174.85,133) and (170.5,128.65) .. (170.5,123.29) -- cycle ;

\path  [shading=_xo1jyet8b,_07y1vsxxi] (300,55.9) .. controls (300,50.71) and (304.21,46.5) .. (309.4,46.5) -- (432.6,46.5) .. controls (437.79,46.5) and (442,50.71) .. (442,55.9) -- (442,84.1) .. controls (442,89.29) and (437.79,93.5) .. (432.6,93.5) -- (309.4,93.5) .. controls (304.21,93.5) and (300,89.29) .. (300,84.1) -- cycle ; 
 \draw   (300,55.9) .. controls (300,50.71) and (304.21,46.5) .. (309.4,46.5) -- (432.6,46.5) .. controls (437.79,46.5) and (442,50.71) .. (442,55.9) -- (442,84.1) .. controls (442,89.29) and (437.79,93.5) .. (432.6,93.5) -- (309.4,93.5) .. controls (304.21,93.5) and (300,89.29) .. (300,84.1) -- cycle ; 

\draw    (324,70.5) -- (345,70.5) ;
\draw [shift={(347,70.5)}, rotate = 180] [color={rgb, 255:red, 0; green, 0; blue, 0 }  ][line width=0.75]    (10.93,-3.29) .. controls (6.95,-1.4) and (3.31,-0.3) .. (0,0) .. controls (3.31,0.3) and (6.95,1.4) .. (10.93,3.29)   ;
\draw   (309,70.5) .. controls (309,66.36) and (312.36,63) .. (316.5,63) .. controls (320.64,63) and (324,66.36) .. (324,70.5) .. controls (324,74.64) and (320.64,78) .. (316.5,78) .. controls (312.36,78) and (309,74.64) .. (309,70.5) -- cycle ; \draw   (309,70.5) -- (324,70.5) ; \draw   (316.5,63) -- (316.5,78) ;
\draw    (316.5,101) -- (316.5,80) ;
\draw [shift={(316.5,78)}, rotate = 90] [color={rgb, 255:red, 0; green, 0; blue, 0 }  ][line width=0.75]    (10.93,-3.29) .. controls (6.95,-1.4) and (3.31,-0.3) .. (0,0) .. controls (3.31,0.3) and (6.95,1.4) .. (10.93,3.29)   ;
\draw    (316.5,40) -- (316.5,61) ;
\draw [shift={(316.5,63)}, rotate = 270] [color={rgb, 255:red, 0; green, 0; blue, 0 }  ][line width=0.75]    (10.93,-3.29) .. controls (6.95,-1.4) and (3.31,-0.3) .. (0,0) .. controls (3.31,0.3) and (6.95,1.4) .. (10.93,3.29)   ;
\draw    (293.5,40) -- (316.5,40) ;
\draw    (293.5,101) -- (316.5,101) ;
\draw    (133,33) -- (164,33) ;
\draw [shift={(166,33)}, rotate = 180] [color={rgb, 255:red, 0; green, 0; blue, 0 }  ][line width=0.75]    (10.93,-3.29) .. controls (6.95,-1.4) and (3.31,-0.3) .. (0,0) .. controls (3.31,0.3) and (6.95,1.4) .. (10.93,3.29)   ;
\draw    (133,109) -- (164,109) ;
\draw [shift={(166,109)}, rotate = 180] [color={rgb, 255:red, 0; green, 0; blue, 0 }  ][line width=0.75]    (10.93,-3.29) .. controls (6.95,-1.4) and (3.31,-0.3) .. (0,0) .. controls (3.31,0.3) and (6.95,1.4) .. (10.93,3.29)   ;
\draw    (447,70.5) -- (468,70.5) ;
\draw [shift={(470,70.5)}, rotate = 180] [color={rgb, 255:red, 0; green, 0; blue, 0 }  ][line width=0.75]    (10.93,-3.29) .. controls (6.95,-1.4) and (3.31,-0.3) .. (0,0) .. controls (3.31,0.3) and (6.95,1.4) .. (10.93,3.29)   ;

\draw (157,13.28) node [anchor=north west][inner sep=0.75pt]   [align=left] {\begin{minipage}[lt]{73.58pt}\setlength\topsep{0pt}
\begin{center}
Communication\\Precoder
\end{center}

\end{minipage}};
\draw (185,89.72) node [anchor=north west][inner sep=0.75pt]   [align=left] {\begin{minipage}[lt]{44.11pt}\setlength\topsep{0pt}
\begin{center}
Radar\\Precoder
\end{center}

\end{minipage}};
\draw (-1,13.28) node [anchor=north west][inner sep=0.75pt]   [align=left] {\begin{minipage}[lt]{73.58pt}\setlength\topsep{0pt}
\begin{center}
Communication\\Waveform
\end{center}

\end{minipage}};
\draw (20,89.72) node [anchor=north west][inner sep=0.75pt]   [align=left] {\begin{minipage}[lt]{48.82pt}\setlength\topsep{0pt}
\begin{center}
Radar \\Waveform
\end{center}

\end{minipage}};
\draw (141,10) node [anchor=north west][inner sep=0.75pt]   [align=left] {$s_{\hat{k}}$};
\draw (298,10) node [anchor=north west][inner sep=0.75pt]   [align=left] {$\mathbf{x}_{\hat{k}}=\mathbf{w}_{\hat{k}}\,s_{\hat{k}}$};
\draw (139,115) node [anchor=north west][inner sep=0.75pt]   [align=left] {$s_0$};
\draw (301,115) node [anchor=north west][inner sep=0.75pt]   [align=left] {$\mathbf{x}_0=\mathbf{w}_0\,s_0$};
\draw (453,44) node [anchor=north west][inner sep=0.75pt]   [align=left] {$\mathbf{x}$};
\draw (333,51) node [anchor=north west][inner sep=0.75pt]   [align=left] {\begin{minipage}[lt]{59.03pt}\setlength\topsep{0pt}
\begin{center}
Transmitted \\Signal
\end{center}

\end{minipage}};

\end{tikzpicture}
\vspace{0.1cm}
    \caption{Beamforming transmission for both communication and sensing.}  \label{fig:precoding}
\end{figure}
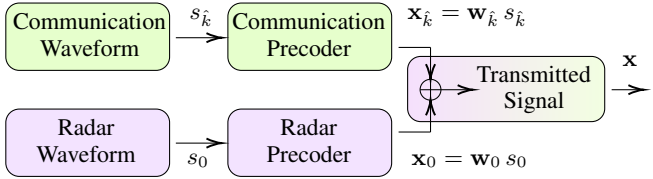

As highlighted in Fig.~\ref{fig:precoding}, we account for a \ac{DFRC} system with separate waveforms for implementation simplicity rather than methods based on dual-function waveforms. With that said, let $\mathbf{s}(\instant)$ be a $(K+1) \times 1$ vector containing the transmitted waveforms as
\begin{align}\label{eq:txsignal}
    &\mathbf{s}(\instant)=[ \underbrace{s_0(\instant)}_{\text{Sensing}},  \underbrace{s_1(\instant), \ldots,s_{\bsk}(\instant), \ldots, s_{K}(n)}_{\text{Communication}}]^T,
\end{align}
where $s_{\bsk}(\instant)$ represents the transmitted signal from the \ac{UAV} to the $\bsk$-th \ac{GBS} at time slot $\instant$, whereas $s_0(\instant)$ represents the transmitted signal for sensing during the same time slot.

After the \ac{GBS} selection, the \ac{UAV} is attached to a single \ac{GBS}, and \eqref{eq:txsignal} reduces to
\begin{align}\label{eq:txsignal2}
    &\mathbf{s}(\instant)=[ s_0(\instant),  s_{\bskn}(\instant)]^T \in \mathcal{C}^{2\times 1},
\end{align}
where $\bskn$ is the chosen \ac{GBS} at time instant $\instant$. 

We consider the \ac{UAV} applying beamforming techniques for communications and sensing. We denote with $\mathbf{W}(\instant) \in \mathcal{C}^{M \times K+1}$ the beamforming matrix given by
\begin{align}
    &\mathbf{W}(\instant)=[ \underbrace{\mathbf{w}_0(\instant)}_{\text{Sensing}},  \underbrace{\mathbf{w}_{\bskn}(\instant)}_{\text{Comm.}}] \in \mathcal{C}^{M \times 2}
\end{align}
with $\mathbf{w}_{\bskn}(\instant) \in \mathcal{C}^{M \times 1}$ representing the corresponding transmit beamforming vector to the \ac{GBS}, whereas $\mathbf{w}_0(\instant)$ is the beamforming vector used for sensing. The generic beamforming vector, for $i \in \mathcal{I} = \{0, \bskn \}$, is given by
\begin{align}
    &\mathbf{w}_i(\instant)=\left[ e^{\jmath \varphi_{i,0}}, \ldots, e^{\jmath \varphi_{i,\ant}}, \ldots, e^{\jmath \varphi_{i,M-1}} \right]^T \in  \mathcal{C}^{M \times 1},
\end{align}
where $\varphi_{i,\ant}$ is the phase at the $\ant$-th antenna of the \ac{UAV} array and $M$ is the total number of antennas.

Accordingly, the transmitted signal from the \ac{UAV} to the $i$th destination, GBS or target, is
\begin{align}
  & \mathbf{x}_i(\instant)= \mathbf{w}_i(\instant) s_i(\instant) \in \mathcal{C}^{M \times 1}, \quad i \in \mathcal{I} = \left\{0, \bskn \right\},
\end{align}
and the total transmitted signal per antenna is, as for \figref{fig:precoding}
\begin{align}
  & \mathbf{x}(\instant)= \mathbf{w}_0(\instant) s_0(\instant) + \mathbf{w}_{\bskn}(\instant) s_{\bskn}(\instant). 
\end{align}
The transmitted power is thus given by
\begin{align}
&\mathbb{E}\left[ \mathbf{x}(\instant)^H \mathbf{x}(\instant) \right] = \mathbf{w}_0(\instant)^H  \mathbb{E}\left[s_0(\instant)^H  s_0(\instant)\right]\mathbf{w}_0(\instant) \nonumber \\
   &\qquad + \mathbf{w}_{\bskn}(\instant)^H  \mathbb{E}\left[s_{\bskn}(\instant)^H 
 s_{\bskn}(\instant)\right]\mathbf{w}_{\bskn}(\instant), 
\end{align}
where $\left( \cdot \right)^H$ is the Hermitian operator and where we have assumed $\mathbb{E}\left[s_{0}(\instant)^H s_{\bskn}(\instant)\right]=0$ (orthogonality) \cite{hua2023optimal_TVT}. Then, by further assuming 
\begin{align}
    &\mathbb{E}\left[s_{\bskn}(\instant)^H s_{\bskn}(\instant)\right]=1 \nonumber, \\
    & \mathbb{E}\left[s_{0}(\instant)^H s_{0}(\instant)\right]=1,
\end{align}
we can write
\begin{align}\label{eq:meanx}
\mathbb{E}\left[\mathbf{x}(\instant)^H\,  \mathbf{x}(\instant) \right] 
& =  \lVert \mathbf{w}_0(\instant)\rVert^2 + \lVert \mathbf{w}_{\bskn}(\instant) \rVert^2.
\end{align}
Hence, according to \eqref{eq:meanx}, the constraint of the maximum available power can be expressed as
\begin{equation}\label{eq:PowerConstraint}
    \lVert \mathbf{w}_0(\instant)\rVert^2 + \lVert \mathbf{w}_{\bskn}(\instant) \rVert^2 \leq P_{\text{max}}.
\end{equation}

\subsection{Air-to-Ground Channel Model} \label{ChannelModel}
\ac{THz} communication enables a high achievable rate thanks to the available bandwidth. However, the \ac{THz} spectrum, lying in between microwaves and optical regions, suffers from high propagation losses and is sensitive to \ac{NLoS} conditions.
\acp{UAV} have the capability to have a privileged position in the 3D space and, thus, have a better chance of creating a \ac{LoS} link with the intended target and \ac{GBS}. 
Deploying high directive antennas with high directivity through large arrays on the \ac{UAV} effectively mitigates the \ac{THz} propagation losses \cite{azari2022:THZ_UAV_IEEECOMM_MAGAZ}.

In this section, we characterize the wireless channel between a possible destination, either the selected \acp{GBS} or the target, and the \ac{UAV} by both deterministic large-scale path loss and random small-scale fading. As widely adopted in works dealing with \ac{UAV} trajectory \cite{zhang2018joint}, we use a probabilistic path loss model, where the \ac{LoS} and \ac{NLoS} links are considered separately with different path loss exponents and probabilities. At time slot $\instant$, the path loss between the \ac{UAV} and the $i$th destination can be written as 
\begin{align}
    \mathsf{PL}_{i}(\instant) =& \mathbb{P}_{\mathrm{LoS},i}(\instant)\,\mathsf{PL}_{\mathrm{LoS},i}(\instant)+\mathbb{P}_{\mathrm{NLoS},i}(\instant)\,\mathsf{PL}_{\mathrm{NLoS},i}(\instant),
\end{align}
where $\mathbb{P}_{\mathrm{LoS},i}$ and $\mathbb{P}_{\mathrm{NLoS},i}$ are the probabilistic occurrences of the \ac{LoS} and \ac{NLoS} links, related to the $i$th destination, and characterized by a path loss equal to $\mathsf{PL}_{\mathrm{LoS},i}(\instant)$ and $\mathsf{PL}_{\mathrm{NLoS},i}(\instant)$, respectively.
We model the \ac{LoS} probability between the \ac{UAV}, of height $\zn$, and the $i$th destination, of height $z_i$, as follows
\begin{align} \label{eq:Los_probab_approx}
&\mathbb{P}_{\mathrm{NLoS}, i}(\instant) =\nonumber \\
&-\kappa_1 \exp\bigg\{-\kappa_2 \cdot \operatorname{atan} \bigg(\frac{\zn-z_{i}}{\di(\instant)\, \sin\left( \thetai(\instant)\right)}\bigg)\bigg\}+\kappa_3,
\end{align}
where $\di(\instant)= \lVert \qn-\ui \rVert_2$ is the distance between the \ac{UAV} and the destination whereas different values of $\kappa_1$, $\kappa_2$ and $\kappa_3$ lead to different propagation scenarios \cite{fontanesi2020outage}.

We assume the signal through the THz band is affected by both free space loss and molecular absorption \cite{chen2022:IEEEtutorial_THz}. Then, to model the path loss of the \ac{LoS}, we need to discriminate between two configurations, that is, $i=0$ (sensing) and $i=\bskn$ (communication).

In the sensing case, that is, when $i=0$, we can exploit the radar range equation to model the path loss as \cite{balanis2015antenna}
\begin{align}
\mathsf{PL}_{\mathrm{LoS}, 0}(\instant) =  \frac{\lambda^2}{(4\pi)^3 \, d_0^4(\instant)}\rho_{\mathrm{T}} \cdot \exp\left(4 K_{\fc}d_0(\instant)\right),
\end{align}
where the fourth power of the distance encapsulates the effect of the wave traveling the round-trip-channel between the target and the \ac{UAV}, $d_0(n)$ is the \ac{UAV}-target distance at time instant $n$, and $\rho_{\mathrm{T}}$ represents the target radar cross-section. Note that $K_{\fc}$ represents the overall absorption coefficient
of the transmission medium at the central (subcarrier) frequency $\fc$ since, for THz-band signals, there is the molecular absorption caused by water vapor and other gases that increases the path loss. Thus, in the radar range equation, we have accounted twice for such an effect through the term $\exp\left(4 K_{\fc}d_0(\instant)\right)$~\cite{chen2022:IEEEtutorial_THz}.

For the communication case, that is, when $i=\bskn$, again, the main component is dictated by the free space loss, whose path loss can be calculated as \cite{chen2022:IEEEtutorial_THz}
\begin{equation}
\mathsf{PL}_{\mathrm{LoS}, \bskn}(\instant) =  \frac{\lambda^2}{(4\pi \, d_{\bskn}(\instant))^2} \exp\left(2{ K_{\fc}d_{\bskn}(\instant) }\right).
\end{equation}

The respective path loss in NLOS can be defined as \cite{chen2022:IEEEtutorial_THz}
\begin{align}
    \mathsf{PL}_{\mathrm{NLoS}, \bskn}(\instant) &=  \mathsf{PL}_{\mathrm{LoS}, \bskn}(\instant)\, K^2_{\mathrm{N}}  \nonumber \\
    &= \frac{\lambda^2}{\left(4\pi d_{\bskn}(\instant)\right)^2}K^2_{\mathrm{N}} \exp\left(2{ K_{\fc}d_{\bskn}(\instant) }\right),
\end{align}
where $K_{\mathrm{N}}$ is the \ac{NLoS} attenuation-loss coefficient.

\subsection{Received Signal Model}

The steering vector $\mathbf{a}\left( \qn, \ui \right)$ towards the $\dest$th destination can be expressed as \cite{guerra:SingleAnchorTWCOM2018} 
\begin{align}\label{eq:asteer}
\mathbf{a}\left( \qn, \ui \right) = &\big[e^{\jmath 2\, \pi\, \fc \, \tau_0}, \ldots, e^{\jmath 2\, \pi\, \fc\, \tau_{\ant}},  \ldots, e^{\jmath 2\, \pi\, \fc\, \tau_{M-1}} \big],
\end{align}
where $\fc$ is the central frequency where the inter-antenna delay can be formulated as
\begin{align}\label{eq:taum}
   \tau_{\ant}\left(\alpha, \beta, \gamma \right) &= \frac{1}{c}\,\posm \,\vec{i}\left(\thetak, \phik\right) \nonumber \\
   & =\frac{1}{c}\left[ \xm (\alpha, \beta, \gamma)\cos(\phik) \sin(\thetak), \right.\nonumber \\
    &\quad\quad \,\,\ym(\alpha, \beta, \gamma)\sin(\phik) \sin(\thetak), \nonumber \\
    &\quad\quad \,\,\zm (\alpha, \beta, \gamma)\left. \cos(\thetak) \right]. 
\end{align}

As a result, the channel vector between the UAV and the $\dest$th destination can be written as the following $M \times 1$ vector:
 \begin{align}
     \mathbf{h}_{\dest}(\qn, \ui) = \sqrt{\mathsf{PL}_{\dest}(\instant)}\, e^{\jmath 2\, \pi \fc \frac{\di(\instant)}{c}} \, \mathbf{a}\left( \qn, \ui \right),
 \end{align}
and the received signal at the $\dest$th destination can be written as
  \begin{equation}
     y_{\dest}(\instant) = \mathbf{h}_{\dest}^H(\qn, \ui) \mathbf{x}_{\dest}(\instant) + {\nu}_{\dest}(\instant) \in \mathcal{C},
 \end{equation}
 where $\nu_{\dest}(\instant)$ is the additive circular complex white Gaussian noise (AWGN) with variance $\sigma^2$. 

Note that the \ac{TOA} between the $\ant$th antenna of the \ac{UAV} and the $\dest$th destination at time $\instant$ is approximated as
\begin{equation}\label{eq:tau_TOA}
    \tau_{\ant,\dest}\left(\instant \right) \approx \tau_{\ant}+ \frac{\lVert \ui-\qn \rVert}{c},
\end{equation}
where $c$ is the speed of light and $\tau_{\ant}$ is the propagation time from the $\ant$th antenna to the array center located in $\qn$, as reported in Fig.~\ref{fig:geometry}.

Then, by defining the angles between the transmitter (UAV) and the receiver (GBS)  as
\begin{align}
    & \thetai= \operatorname{acos}\left( \frac{\zn-\zi}{\di\left(\instant\right)} \right),\\
    & \phii= \operatorname{atan}\left( \frac{\yn-\yi}{\xn-\xii}\right),
\end{align}
with $\di\left(\instant\right)=\lVert \ui-\qn \rVert$, and the direction vector as
\begin{align}
     \vec{i}\left(\theta, \phi\right) = &\left[ \cos(\phi) \sin(\theta), \right.\nonumber \\
    &\sin(\phi) \sin(\theta), \nonumber \\
    &\left. \cos(\theta) \right].
\end{align}

Note that this \ac{TOA} depends on the antenna array coordinates and the rotational angles (to be optimized).
Indeed, the \ac{DFRC} \ac{UAV} can adapt its beamforming weights and orientation according to communication and sensing needs. In addition, the ground \ac{GBS} association problem previously illustrated is included.

\section{Problem Formulation}\label{ProblemFormulation}
This section illustrates the joint optimization problem for balancing and finding a tradeoff between communications and sensing needs. Before presenting the problem, we separately introduce the metric functions to be optimized for both functionalities.

\paragraph{Communication metric} For enhancing the communications between the \ac{UAV} and the selected \ac{GBS}, we consider the \ac{SINR} at the $\bskn$-th \ac{GBS} formulated as
\begin{align}
\label{eqn_dbl_x}
    \mathsf{SINR} (\mathbf{q}(n),\mathbf{u}_{\bskn}) = \frac{|\mathbf{h}_{\bskn}^H(\mathbf{q}(n),\mathbf{u}_{\bskn})\mathbf{w}_{\bskn}(n)|^2}{\sigma^2 + |\mathbf{h}_{\targ}^H(\mathbf{q}(n),\mathbf{u}_{\targ})\mathbf{w}_{0}(n)|^2}.
\end{align}
We have considered the worst-case scenario where the receiver cannot cancel the radar signals' interference before decoding its desirable information signal \cite{hua2023optimal_TVT}.

\paragraph{Sensing metric} A key performance metric adopted in literature for radar signal design is the transmit beam-pattern \cite{fuhrmann2008:transmitBeampattern_TransAreosp,liu2020:jointTXBeam_MIMO}.
The transmit beam pattern describes the transmit signal power at a generic focal point pointed towards the potential target in the interested area.
Without prior knowledge of the target location, the transmit signal power is pointed at any value in the range $ [-\pi/2, \pi/2]$ to perform a target search in any direction.
Here, we assume the target is fixed in one position and that the position is known a priori. In the following, we note the estimated position of the target as $\mathbf{\hat{u}}_0$. 
 
In this case, starting from \cite{liu2020:jointTXBeam_MIMO}, we can formalize the transmit beam-pattern gain towards location $\poss$ as
\begin{align}\label{eq:BeampatternGain}
 B(\qn, \poss)& =   \nonumber \\
&(\mathbf{a}\left( \qn, \poss \right))^H (R_d + \lVert \mathbf{w}_{\bskn}(\instant) \rVert^2 )\mathbf{a}\left( \qn, \poss \right). 
\end{align}
We highlight that on the one hand, $\mathbf{a}(\qn, \poss)$ is a known function of $\theta$ since we initially assume a single static target, such that $j=1, u_1 = (x_1, y_1, z_1)$. On the other hand, the beam pattern $B$ varies along the trajectory because we assume that the angle between the target and the \ac{UAV} changes through time.

Then, let  $B^\star(\mathbf{q}_{\ell}, \ui)$ denote the desired beam pattern, which specifies the desired beamforming weights and directions for the $\mathrm{N}$ UAV positions along the $\ell$th trajectory.

We can then formalize a cost function $C$ that defines the beam-pattern matching error along the $\ell$th trajectory as 
\begin{equation}\label{eq:beampatternError}
    C(\mathbf{q}_{\ell}(n), \mathbf{W}_{\ell}(\instant))=\sum_{\forall \ell,n=1}^N  |{B^\star(\mathbf{q}_{\ell}(n), \poss}) - B(\mathbf{q}_{\ell}(n), \poss)|^2,
\end{equation}
where we added subscript ${\ell}$ to $\mathbf{W}$ to indicate the beamforming matrix associate to the ${\ell}$-trajectory.

We note that, according to \eqref{eq:BeampatternGain}, we would like to choose $R_d$ and $\mathbf{w}_{\bskn}$ under the total transmit power constraint such that the available transmit power is used to maximize the signal power at the locations of interest. 

\begin{figure*}[t!]
\centering
\includegraphics[width=0.75\linewidth]{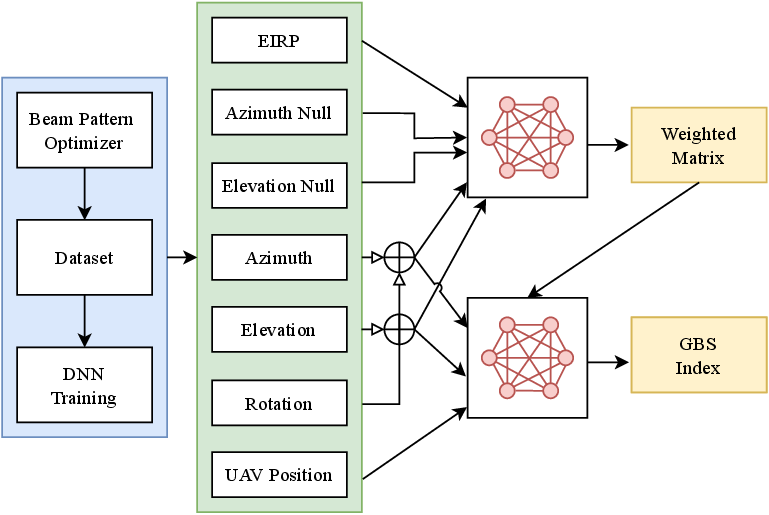}
\vspace{0.1cm}
    \caption{Architecture and inputs/outputs of the proposed \ac{NN} solution.}
    \label{fig:NNSolution}
\end{figure*}

\paragraph{Problem formulation} Our objective is to minimize the beam-pattern gain error, expressed in \eqref{eq:beampatternError} optimizing $R_d$ and $\mathbf{w}_{\bskn}$ together with the optimal \ac{GBS} $\bskn$, subject to a connectivity constraint. 
Thus, our optimization problem can be formulated as 
\begin{subequations}\label{eq:Opt_problem1}
    \begin{align}
  \mathcal{P}:  \min_{\mathbf{o}, \mathbf{W_{\ell}}, \bskn} \quad &  C(\mathbf{q}_{\ell},\mathbf{W}_{\ell}(\instant))\\ \label{eq:cond1_Opt_problem1_discrete}
    \textrm{s.t.} \quad & \zn > z_{\bskn}=h_{\text{BS}},\\
    \label{eq:cond2_Opt_problem1_discrete}
    & \mathbf{q}(1)=\mathbf{q
    }_{\mathrm{I}}, \mathbf{q}(N)=\mathbf{q}_{\mathrm{F}}\\
     \label{eq:cond3_Opt_problem1_discrete} 
     &  R_d(n) + \lVert \mathbf{w}_{\bskn}(\instant) \rVert^2 \leq P_{\text{max}} ,\\ \label{eq:cond5_Opt_problem1_discrete}
    &  \mathsf{SINR}(\qn,\mathbf{u}_{\bskn}) \geq \gamma_{\text{SINR}}.
    \end{align}
    \end{subequations}
where $\mathbf{o} = [\alpha, \beta, \gamma], $ \eqref{eq:cond1_Opt_problem1_discrete} ensures the UAV height respects the local height regulations, with $h_{\text{BS}}$ indicating the minimum required height; \eqref{eq:cond2_Opt_problem1_discrete} guarantees that the \ac{UAV} starts and ends its trajectory at the required locations. Then, 
\eqref{eq:cond3_Opt_problem1_discrete} denotes the power constraint on the \ac{UAV}, with $P_\text{max}$ denoting the maximum tolerable transmitted power. Finally, \eqref{eq:cond5_Opt_problem1_discrete} ensures the minimum connectivity constraint with the \ac{GBS}.

Note that, in general, it is difficult optimally solve problem $\mathcal{P}$ for the following reasons:
\begin{itemize}
    \item The cost function is not convex. To make it convex, an auxiliary variable should be introduced \cite{fuhrmann2008:transmitBeampattern_TransAreosp}. However, this would add another constraint to the problem, increasing the complexity.
    \item Even if the objective cost function can be turned into a linear function, modeling the end-to-end channel of \ac{UAV}-\ac{BS} to derive \eqref{eq:cond5_Opt_problem1_discrete} requires accurate channel modeling and perfect global channel information. This might be complicated or even impractical for the THz frequency band under consideration. 
    \item Lastly, due to the sharp \ac{LoS}/\ac{NLoS} transitions, the convex shaped \ac{GBS} coverage property may not hold \cite{chen:jointBSAssociaionTraj_IoT2021}
\end{itemize}
Consequently, in the next section, we propose a double DNN approach.

\section{Proposed DNN Approach}
\label{sec:proposed}
We propose a data-driven approach to develop a learning-based beamforming scheme to address problem $\mathcal{P}$. 
To this purpose, we note that \eqref{eq:Opt_problem1} can be considered as a beamforming design problem for each \ac{UAV}'s position along the trajectory $\mathbf{q}_{\ell}$. 
Notably, the control of the constraints in \eqref{eq:Opt_problem1} and of the \ac{UAV} rotation while finding the beamforming matrix is a complex task and requires synthesizing real-world real-time beamforming patterns for any possible \ac{UAV} trajectory point.

Currently developed ad-hoc tools, like beam-pattern optimizers \cite{j_vazquez2023_surrogate}, allow to manage for a single point at a time the different requirements of the beamforming applied at the \ac{DFRC} \ac{UAV}. 
For this reason, we compute training beamforming matrices for a different set of UAV trajectories that satisfy conditions \eqref{eq:cond1_Opt_problem1_discrete}, \eqref{eq:cond2_Opt_problem1_discrete}, \eqref{eq:cond3_Opt_problem1_discrete} using a beam-pattern optimizer.
Then, we develop a \ac{DNN} that synthesizes real-time realistic antenna beam patterns considering the radiation pattern's possible rotation.

The developed predictive beamforming framework is illustrated in \figref{fig:NNSolution} and consists of two phases: $a)$ training with a Beam-Pattern Optimizer and $b)$ a double \ac{DNN} solution.
We will describe in what follows the two phases separately.

\subsection{Beam-Pattern Optimizer for Training Data Generation}\label{SurrogateOptimizer}

During the first phase $a)$, a beamforming matrix for a different set of \ac{UAV} trajectories is created such that it satisfies conditions \eqref{eq:cond1_Opt_problem1_discrete}, \eqref{eq:cond2_Opt_problem1_discrete}, \eqref{eq:cond3_Opt_problem1_discrete}. 

We consider a realistic antenna and a realistic number of antenna elements to perform the communication and sensing task and compute the inherent footprint. 
Accordingly, we takes under consideration a patch antenna that generates a cardioid radiation pattern and a maximum number of $M=100$ elements.
The direction angles, measured in azimuth and elevation, are defined geometrically by the positions of the \ac{UAV}, of the associated \ac{GBS}, and the target.
According to the aforementioned considerations, the nulling operation has to be performed in the directions of the non-attached \acp{GBS}.
\begin{algorithm}[t!]
\SetAlgoLined
\textbf{Input:}Initial UAV location\;  
\hspace{1cm}EIRP$^*$, EIRP per beam,\;
\hspace{1cm}$\sllmin$, SLL minimum per beam\;
\hspace{1cm} ($\thetan, \phin$), Null position per beam\;
\textbf{Output:}$\wpq$, Weight matrix based on previous inputs\;
\textbf{Data:} Set of possible configurations on UAV considering system constraints\;
  $F_1=$ initial value\;
   \textbf{Data:} Beam-Pattern Optimizer\;
   \If{$counter < counter_{max}$}
    {
  \textbf{Compute:}  $\wpq$\\
  \textbf{Compute:} radiation pattern, $\sllec$, $\sllac$ and $\eirpc$\;
  \textbf{Compute:} $F(Z_1(\wpq) + Z_2(\wpq))$\;
                  \eIf{$F<\eta$}
                  {
                  $\wpq^\star = \wpq$\;
                  saves the optimal matrix\;
                  
                   \textbf{break}
                  
                  }
                  {
                 $counter \gets counter + 1$\;
                  
                  } 
                  
        \textbf{Optimize:} $\row$, $\col$, $\slla$, $\slle$, and $\ppe$\;  
  }
  Save the optimal matrix $\wpq$\;
 \caption{Beam-Pattern Optimizer}
 \label{alg:SurrogateOptimizer}
\end{algorithm}

The beam-pattern optimizer operates by activating or deactivating the number of active rows ($\ant_x$) and columns ($\ant_z$) in the array and finding the proper tapering by selecting the adequate \ac{SLL} in both planes to generate the required $\thetat$ in the elevation ($\thetae$) and azimuth planes ($\thetaa$), and computing the required power per element to address the necessary \ac{EIRP}.
For each position of the \ac{UAV}, we have to compute two array weights vectors: (i) one dedicated to the sensing operation and (ii) one dedicated to the communication.

According to the previous considerations, the cost function is composed of the sum of two sub-objectives as follows 
\begin{equation}
\label{eq:CostFunction}
\wpq^\star=\min_{\vspace{1cm} \wpq}     \quad            Z_1(\wpq)+Z_2(\wpq),
\end{equation}
where $\wpq$ indicates the $gv$-th element of the beamforming matrix, $Z_1(\wpq)$ and $Z_2(\wpq)$ are given by

\begin{equation}
\label{eq:Eq2b}
    \left\{
    \begin{aligned}
         Z_1 (\wpq)= &\Bigg(\frac{|\sllac(\wpq)-\sllad|}{\sllad} \\
        & +\frac{|\sllec(\wpq)-\slled|}{\slled}\Bigg) \cdot k_1\\
          Z_2 (\wpq)= &\frac{\mathrm{EIRP}(\wpq)-\mathrm{EIRP}^*}{\mathrm{EIRP}^*}\cdot k_2 \,.\\
    \end{aligned}
    \right.
\end{equation}
Note that $Z_1(\wpq)$ computes the error between the minimum desired \ac{SLL} in both planes, that is, $\sllad, \slled$ compared with their computed counterparts $\sllac, \sllec$ per beam. $Z_2(\wpq)$ reports the error between the desired $\mathrm{EIRP}^*$ and the calculated $\mathrm{EIRP}$ per beam. Each of those terms has a weighting factor $k_1$ and $k_2$ that will add additional importance to their calculation.

\begin{table}[t!]
 \begin{center}
    \footnotesize 
    \begin{tabular}{|l|p{0.2\columnwidth}|p{0.2\columnwidth}|p{0.2\columnwidth}|p{0.2\columnwidth}|}
        \hline
        \emph{Parameters} & \emph{Value}  & \emph{Parameters} & \emph{Value} \\
        \hline
        Area size & $2.25\,\mathrm{km}^2$ & BS height & $2\, \mathrm{m}$ \\
        Number of BS K & $5$ & max EIRP & $37\, \mathrm{dBm}$ \\
        \ac{UAV} height & $100\, \mathrm{m}$ & Bandwidth &  $100\, \mathrm{MHz}$ \\
        \ac{UAV} speed & $10 \mathrm{m}$  & Frequency & $0.3\, \mathrm{THz}$ \\
        $\delta_t$  &  $1\, \mathrm{sec}$ & Noise Power &  $-110\, \mathrm{dBm}$\\
        SNR threshold & $0.3\, \mathrm{dB}$ & \ac{UAV} training trajectories & $100$\\
        \hline
    \end{tabular}
    \end{center}
     \caption{\footnotesize{Simulation Parameters}}\label{tab:SimParams}
\end{table}
\begin{table}[]
\begin{center}
\begin{tabular}{ccccccc}
\hline
\multicolumn{1}{|c|}{} & \multicolumn{1}{c|}{\begin{tabular}[c]{@{}c@{}}SLL\\ (dB)\end{tabular}} & \multicolumn{1}{c|}{\begin{tabular}[c]{@{}c@{}}EIRP\\ (dBm)\end{tabular}} & \multicolumn{1}{c|}{\begin{tabular}[c]{@{}c@{}}$\phi_0$\\ $\theta_0$\end{tabular}} & \multicolumn{1}{c|}{\begin{tabular}[c]{@{}c@{}}$\phi_{\text{target}}$\\ $\theta_{\text{target}}$\end{tabular}} & \multicolumn{1}{c|}{\begin{tabular}[c]{@{}c@{}}$\phi_{\text{Null},1}$\\ $\theta_{\text{Null},1}$\end{tabular}} & \multicolumn{1}{c|}{\begin{tabular}[c]{@{}c@{}}$\phi_{\text{Null},2}$\\ $\theta_{\text{Null},2}$\end{tabular}} \\ \hline
\multicolumn{1}{|c|}{\# 1} & \multicolumn{1}{c|}{\textgreater{}15} & \multicolumn{1}{c|}{15.38} & \multicolumn{1}{c|}{\begin{tabular}[c]{@{}c@{}}$43.6^\circ$\\ $16.2^\circ$\end{tabular}} & \multicolumn{1}{c|}{\begin{tabular}[c]{@{}c@{}}$48.54^\circ$\\ $26.16^\circ$\end{tabular}} & \multicolumn{1}{c|}{\begin{tabular}[c]{@{}c@{}}$12.61^\circ$\\ $42.63^\circ$\end{tabular}} & \multicolumn{1}{c|}{\begin{tabular}[c]{@{}c@{}}$76.64^\circ$\\ $42.4^\circ$\end{tabular}} \\ \hline
\multicolumn{1}{|c|}{\# 2} & \multicolumn{1}{c|}{\textgreater{}23} & \multicolumn{1}{c|}{18} & \multicolumn{1}{c|}{\begin{tabular}[c]{@{}c@{}}$-58.4^\circ$\\ $8.4^\circ$\end{tabular}} & \multicolumn{1}{c|}{\begin{tabular}[c]{@{}c@{}}$8.84^\circ$\\ $11.26^\circ$\end{tabular}} & \multicolumn{1}{c|}{\begin{tabular}[c]{@{}c@{}}$57.09^\circ$\\ $25.47^\circ$\end{tabular}} & \multicolumn{1}{c|}{\begin{tabular}[c]{@{}c@{}}$33.85^\circ$\\ $27.30^\circ$\end{tabular}} \\ \hline
\end{tabular}
\end{center}
\caption{Input parameters for \ac{UAV} trajectory point 1 (\# 1) and 2 (\# 2).}
\label{Table:InputParameters}
\end{table}
\begin{table}[]
\begin{center}
\begin{tabular}{cccc}
\hline
\multicolumn{1}{|c|}{} & \multicolumn{1}{c|}{\begin{tabular}[c]{@{}c@{}}SLL\\ (dB)\end{tabular}} & \multicolumn{1}{c|}{\begin{tabular}[c]{@{}c@{}}EIRP\\ (dBm)\end{tabular}} & \multicolumn{1}{c|}{Active elements} \\ \hline
\multicolumn{1}{|c|}{\# 1} & \multicolumn{1}{c|}{30.2} & \multicolumn{1}{c|}{15.3801} & \multicolumn{1}{c|}{100} \\ \hline
\multicolumn{1}{|c|}{\# 2} & \multicolumn{1}{c|}{25.6} & \multicolumn{1}{c|}{18.001} & \multicolumn{1}{c|}{100} \\ \hline
\end{tabular}
\end{center}
\caption{Output parameters of the radiation pattern presented in \figref{fig:Radpatt_close} for \ac{UAV} Trajectory Point 1 (\# 1) and 2 (\# 2).}
\label{Table:OutputRadPatt}
\end{table}

\begin{figure*}[!htb]
\minipage{0.32\textwidth}
  \includegraphics[width=\linewidth]{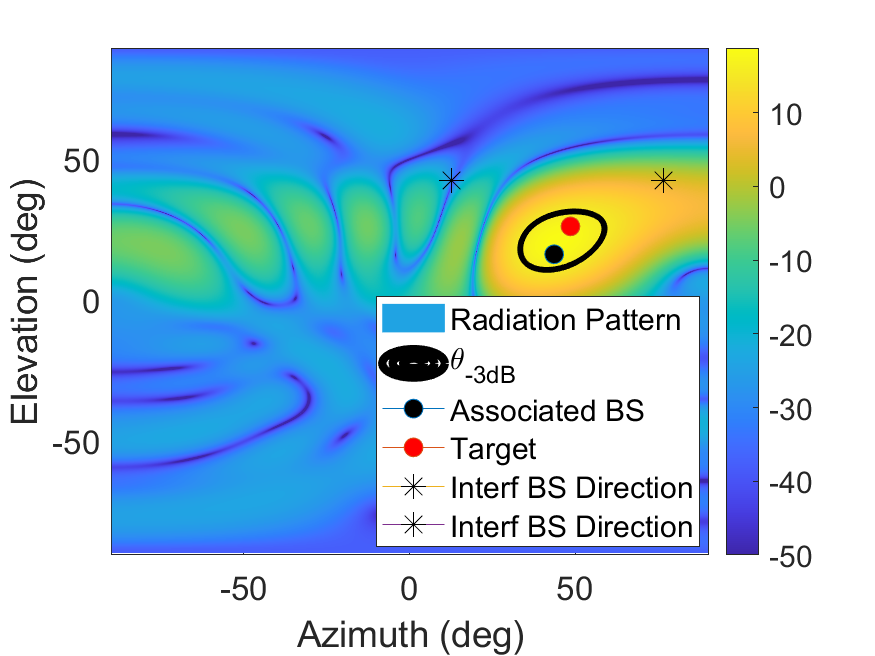}
  \subcaption{ISAC}\label{fig:Fig.RadPatt1}
\endminipage\hfill
\minipage{0.32\textwidth}
  \includegraphics[width=\linewidth]{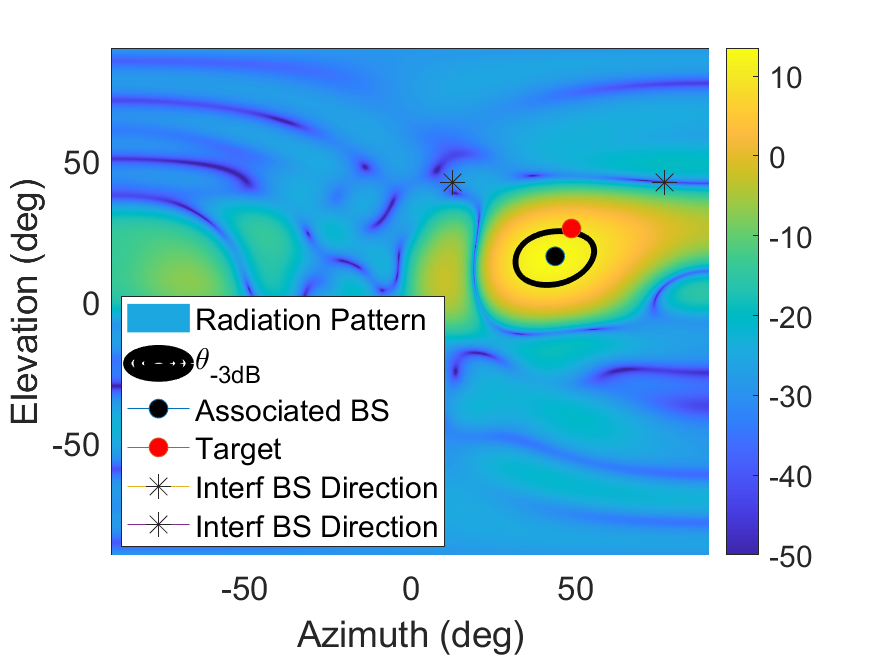}
  \subcaption{Communication}\label{fig:Fig.RadPatt2}
\endminipage\hfill
\minipage{0.32\textwidth}%
  \includegraphics[width=\linewidth]{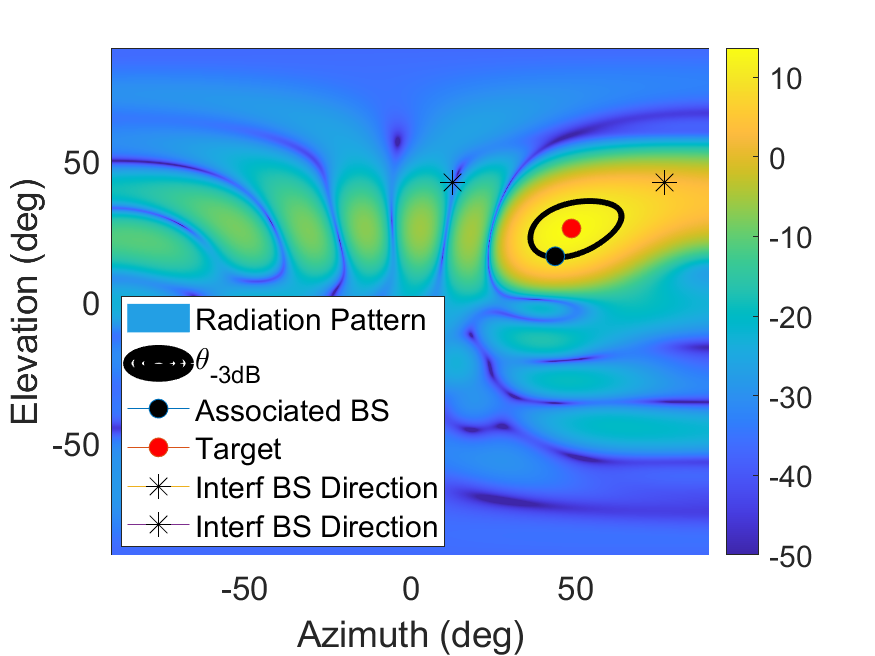}
  \subcaption{Sensing}\label{fig:Fig.RadPatt3}
\endminipage
\caption{Radiation patterns under different requirements in a case where the sensing and the associated BS have close directions: sensing, communication and ISAC. The unit of the colormap is db. The red spot identifies the target area, the black one the position in azimuth and elevation of the associated ground BS while the black stars '*' the non-associated ground BS where to suppress the interference.}
        \label{fig:Radpatt_close}
\end{figure*}

\begin{figure*}[!htb]
\minipage{0.32\textwidth}
  \includegraphics[width=\linewidth]{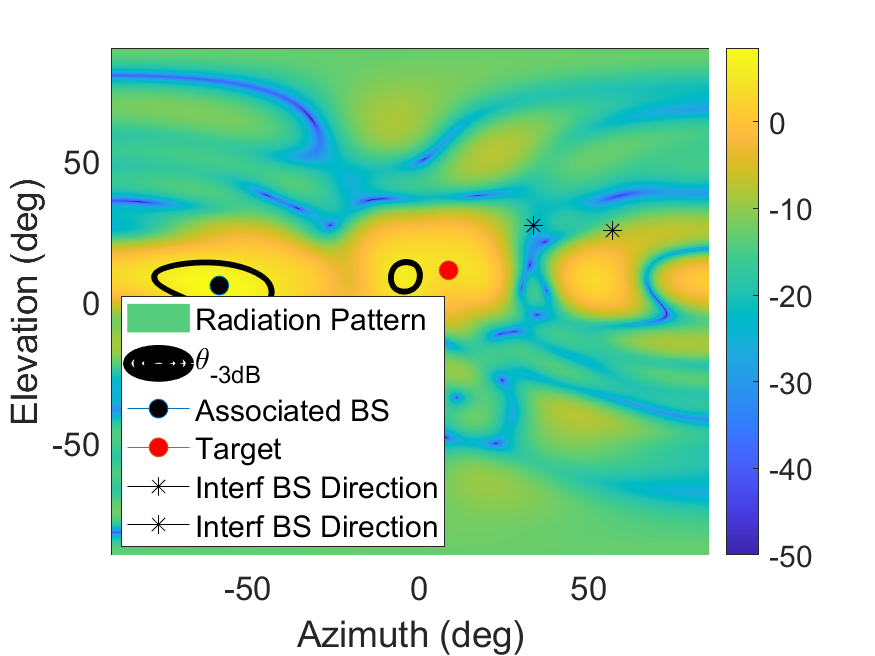}
  \subcaption{ISAC}\label{fig:FigFar.RadPatt1}
\endminipage\hfill
\minipage{0.32\textwidth}
  \includegraphics[width=\linewidth]{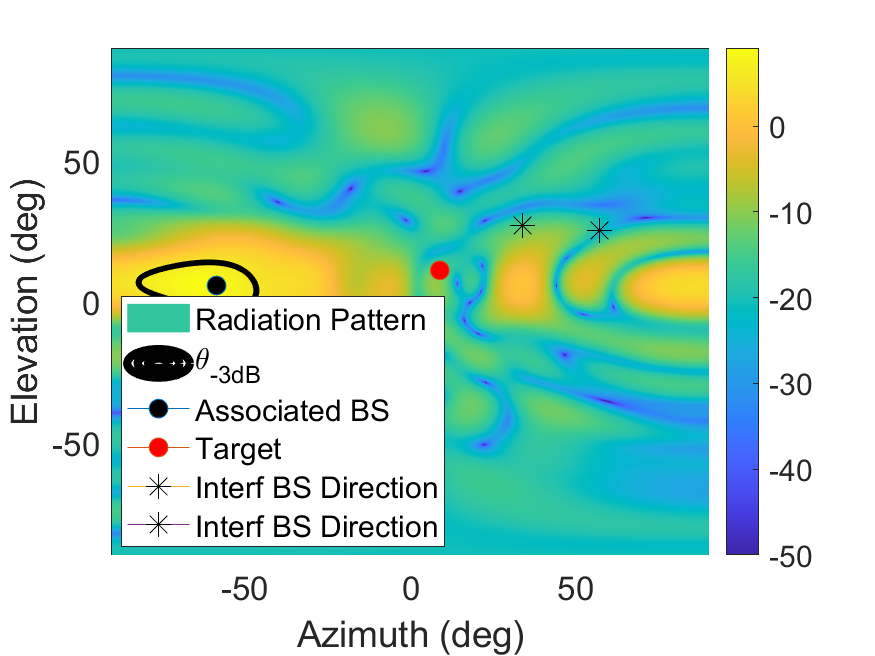}
  \subcaption{Communication}\label{fig:FigFar.RadPatt2}
\endminipage\hfill
\minipage{0.32\textwidth}%
  \includegraphics[width=\linewidth]{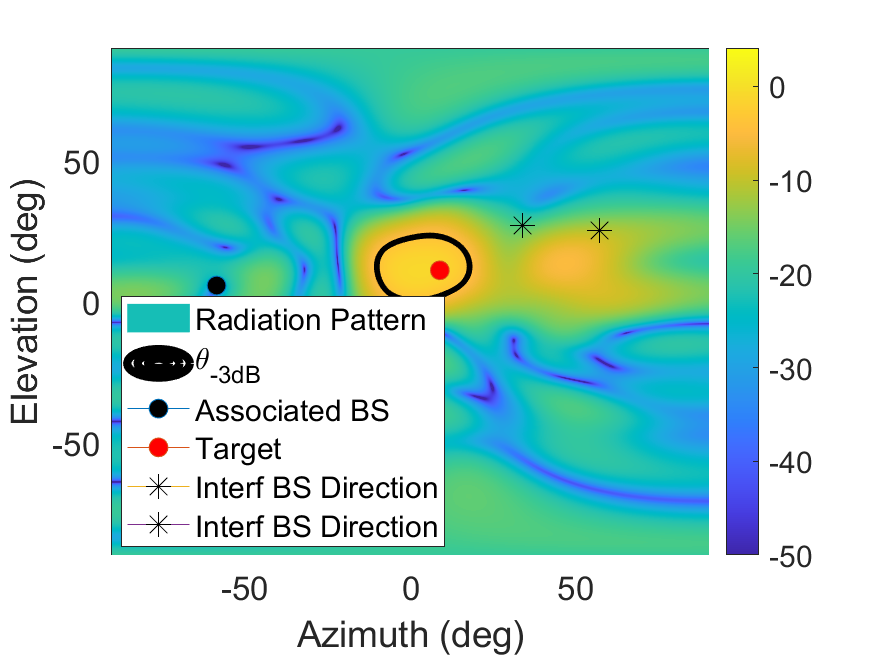}
  \subcaption{Sensing}\label{fig:FigFar.RadPatt3}
\endminipage
\caption{Radiation patterns under different requirements in a case where the sensing and the associated BS have distant directions: sensing, communication and ISAC. The unit of the colormap is db. The red spot identifies the target area, the black one the position in azimuth and elevation of the associated ground BS while the black stars '*' the non-associated ground BS where to suppress the interference.}
        \label{fig:Radpatt_far}
\end{figure*}

With that said, the algorithm operates as follows. First, it requires the azimuth and elevation coordinates of the center of the beam, the tolerable \ac{SLL} range delimited by the minimum ($\sllmin$) values, the required EIRP ($\eirpd$), and the null position ($\thetan, \phin$). Then, the progressive phase shift, nulling, and tapering based on Chebyshev amplitude control are calculated and given as an initial weight matrix $\wpq$ to the optimization algorithm with all the active elements and an initial power per element $\ppe$. Later, the algorithm calculates, for each iteration, the radiation pattern principal cuts per beam, and extracts the \ac{SLL}, nulling, and \ac{EIRP} for both cuts. Based on the previously extracted parameters, the algorithm calculates the cost function $F$, which, if it is lower than the minimum threshold, $\eta$, then the algorithm stops, and the optimum weight matrix $\wpq^\star=\wpq$ will be the output. Note that the threshold is selected to have a small error between the optimized and required values. On the other hand, if the cost function is above the threshold, the algorithm increases the counter and searches for a suitable active number of rows $\row$, columns $\col$, a Chebyshev taper based on the \ac{SLL} admissible range, and power per element $\ppe$ and repeat the previous calculation until it found the optimal weight matrix.

The main steps are reported in Algorithm~\ref{alg:SurrogateOptimizer}.

\subsection{Proposed \ac{NN} Solution}
Problem \ref{eq:Opt_problem1} is highly non-convex, includes a \ac{GBS} association procedure not considered in the beampattern optimizer, and requires running a new simulation every time the \ac{UAV} moves.
Moreover, the \ac{UAV} might take many trajectories and scan different \acp{GBS} associations.
During the training process and the different \ac{GBS} association along its trajectories, conditions \eqref{eq:cond3_Opt_problem1_discrete} and \eqref{eq:cond5_Opt_problem1_discrete} might be satisfied for the single communication or sensing task, but failed when considered jointly.
Thus, the weight recalculations via a beam optimizer for each UAV position and GBS association would lead to lengthy procedures with high computational resources and power consumption. This is unsuitable due to the need for fast response of the \ac{UAV}.

In this sense, phase $b)$ comes to help first produce the required beamforming matrix for any point of the \ac{UAV} flying area in a reduced time compared to the optimizer.
Second, we also address the \ac{UAV}-\ac{GBS} association problem described in \ref{BS_Association} with constraints \eqref{eq:cond3_Opt_problem1_discrete} \eqref{eq:cond5_Opt_problem1_discrete}, exploiting the output of the first \ac{NN} to train a \ac{DNN} to learn the optimal mapping from the input features to the \ac{GBS} association decision.

In our approach, the output of the beamformer optimizer is used to train the \ac{DNN} accounting for both the sensing and communication performance.
We exploit the ability of \ac{DNN} to map the input-output of generic scenarios in a model-agnostic way \cite{shlezinger2022:Access_MLOpt}.
The beamforming weights for communication and sensing (\figref{fig:precoding}) are generated via the beampattern optimizer presented in the previous section and then used as input training of \ac{DNN} (\figref{fig:NNSolution}). Once trained, the optimized beamforming weights $\mathbf{W_{\ell}}$ can be obtained from the \ac{DNN}.

Then, we employ a \ac{FFNN} implementation, where we load as input for the communication beamforming weights for each \ac{UAV} position and for all the trajectories, the azimuth, and elevation towards the associated \ac{GBS}, the azimuth and elevation of the interfering \ac{GBS} based on the decided \ac{GBS} association rule, and the EIRP (\figref{fig:NNSolution}). To generate the sensing weights, we load to the \ac{FFNN} with the azimuth and elevation towards the target position and the associated EIRP.

\section{Numerical Evaluation}
\label{sec:sims}
\subsection{Scenario Definition}

This section provides extensive simulation data to validate the proposed framework and evaluate system parameters' impact on sensing and communication performance.
For the simulations, we consider an area of $1.5 \times 1.5\,\mathrm{km}^2$, where five \ac{GBS} are distributed. The starting and final \ac{UAV} path points are set respectively at $[0, 0]\, \mathrm{m}$, $[700, 800]\, \mathrm{m}$, and target location is at $[350,  400]\, \mathrm{m}$. 
For ease of illustration, the flight altitude is assumed to be fixed at 100 m during the path \footnote{Our proposed scheme can be extended to \ac{UAV} paths with altitude variation.}.
We have randomly generated $N_T=100$ different trajectories between the starting and final points.
A sub-THz downlink system is considered, operating at $0.3\,\mathrm{THz}$. The remaining simulation parameters are specified in Table \ref{tab:SimParams}.

\subsection{Evaluation of the Proposed DNN solution}
To validate and demonstrate the capabilities of the antenna pattern optimizer that creates the dataset and the proposed \ac{DNN} solution, let us consider two example trajectory points where an \ac{UAV} phased array generates a radiation pattern with the parameters presented in Table \ref{Table:InputParameters}. The antenna optimizer considers an array antenna to produce a beam synthesized with SLL, pointing angle, EIRP, and nulling as input.
The above generated are computed for each point of the \ac{UAV} trajectories as follows: the pointing angle is the angle pointing to the associated \ac{GBS} or target, the nulling is the direction of the two closest neighboring \ac{GBS}, the EIRP is the minimum EIRP to satisfy constraint \ref{eq:cond5_Opt_problem1_discrete}.
\begin{figure}[t!]
\minipage{0.24\textwidth}
  \includegraphics[width=\linewidth]{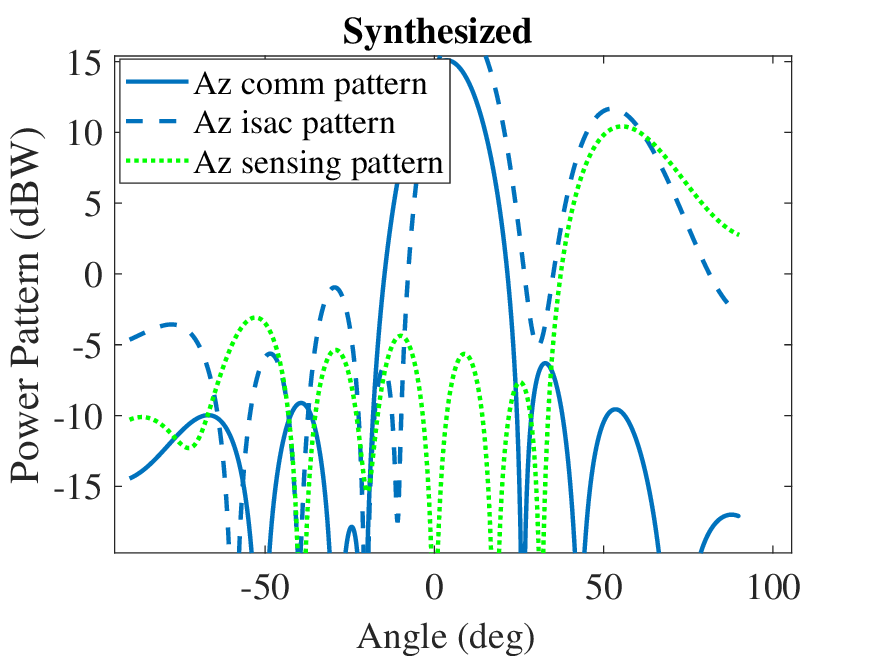}
  \label{fig:AzimuthTransmitBeamPattern}
\endminipage\hfill
\minipage{0.24\textwidth}
  \includegraphics[width=\linewidth]{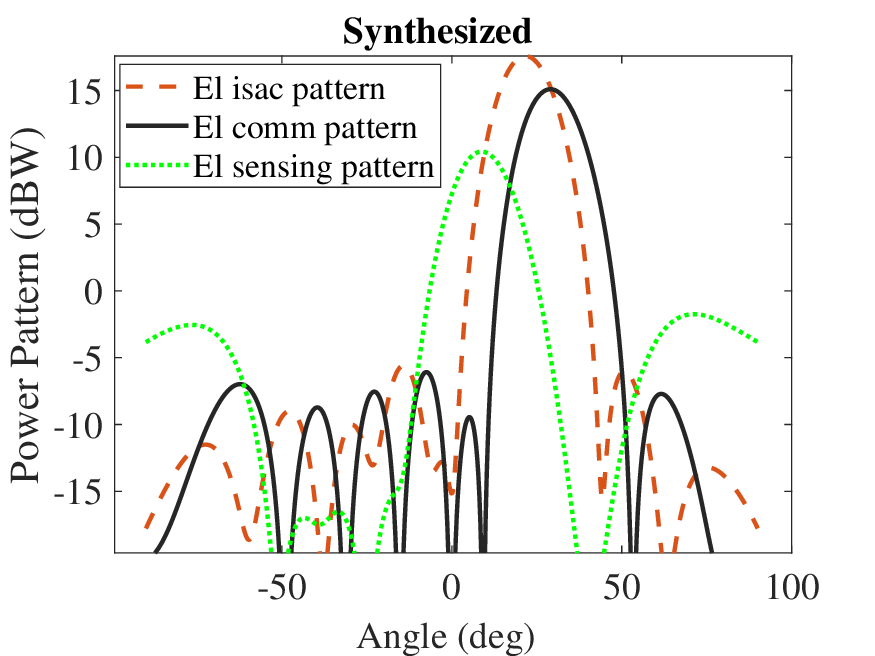}
  \label{fig:ElevationTransmitBeamPattern}
\endminipage\hfill
\caption{Left: synthesized transmit azimuth beampattern profile. Right: synthesized transmit elevation beampattern profile.}
\label{fig:AzimuthElevationTransmitBeamPattern}
\end{figure}
\begin{table}[t]
  \centering

\begin{tabular}{|c|c|c|}
 \hline
 \multicolumn{3}{|c|}{Computational Time} \\
 \hline
 Solution & Training Epochs & Elapsed Time    \\
 \hline
    Beamforming Optimizer & one UAV position &  16 sec \\
    Beamformer NN    &  200 &   47 sec - Training \\
    Beamformer NN    & 1 &  60 msec - Inference  \\
    Association NN    & 200 &  20 sec - Training  \\
    Association NN    & 200 &  30 msec - Inference  \\
 \hline
\end{tabular}
  \caption{Computation time for the DNN proposed solutions}
\label{table:TrainingTime}
\end{table}
The generated $M$ proper complex weights $\textbf{w}_m$, with $m\in \{1,...,M\}$, are used to train the beamforming neural network.
The beamforming network has an input layer of seven neurons corresponding to the desired bandwidth, sidelobes, power antenna constraints, one hidden layer ($50$ neurons), and an output layer ($200$ neurons). The proposed architecture task predicts the beamforming weights for each \ac{UAV} position satisfying \eqref{eq:Opt_problem1}.
The inputs are normalized to improve the optimization performance and the training process. We train the \ac{DNN} by using the ADAM optimizer.
We have considered a slow learning rate and a batch size equal to $128$. We have chosen a ratio of $70-30$ for the training and test set.
The structure of the association network is composed of an input layer with four neurons, corresponding to the \ac{UAV} position in $2$D space. Two hidden layers of $64$ and $32$ neurons follow, and finally, the output layer has one neuron corresponding to the BS index decision.

\figref{fig:Radpatt_close} and \figref{fig:Radpatt_far} show the obtained (transmit) \ac{DNN} beampattern gains in space at specifically chosen trajectory points for the sensing-only design, ISAC design, and the communication-only design, respectively. 
For ease of illustration, angle $(0,0)$ and angles to the radar and \ac{GBS} in azimuth and elevation are computed using the reference system in \figref{fig:geometry}.
The results show that the algorithm generates the three radiation patterns addressing the desired EIRP, SLL requirements and pointing to the desired direction of the main beam and the nulls.
For sensing design in \figref{fig:Fig.RadPatt3} and \figref{fig:FigFar.RadPatt3}, the \ac{UAV} is observed to direct its antenna main lobe ($\theta_{\text{3dB}}$) at the center of the sensing area, and the sensing power exactly covers the whole sensing area, thanks to the adequately designed sensing beams in this case. 
\begin{figure}[tp]
\centering
%
%
\definecolor{mycolor1}{rgb}{0.00000,0.44700,0.74100}%
\definecolor{mycolor2}{rgb}{0.52157,0.75686,0.91373}%
\definecolor{mycolor3}{rgb}{0.20392,0.59608,0.85882}%
\definecolor{mycolor4}{rgb}{1.00000,0.27059,0.00000}%
\definecolor{mycolor5}{rgb}{0.92900,0.69400,0.12500}%
\definecolor{mycolor6}{rgb}{0.25882,0.28627,0.28627}%

\begin{tikzpicture}

\begin{axis}[%
width=2.521in,
height=2.566in,
at={(0.758in,0.481in)},
scale only axis,
bar shift auto,
xmin=-0.2,
xmax=8.2,
xtick={1,2,3,4,5,6,7},
xticklabels={{1},{10},{20},{50},{100},{150},{200}},
xlabel style={font=\color{white!15!black},font=\footnotesize},
xlabel={Training Epochs},
ymin=0,
ymax=0.2,
ylabel style={font=\color{white!15!black},font=\footnotesize},
ylabel={BeamPattern Error},
axis background/.style={fill=white},
legend style={font=\footnotesize,legend cell align=left, align=left, draw=white!15!black}
]
\addplot[ybar, bar width=6, fill=mycolor2, draw=mycolor2, area legend] table[row sep=crcr] {%
1	0.18272339\\
2	0.10387403\\
3	0.065812826\\
4	0.029257132\\
5	0.016647426\\
6	0.013958467\\
7	0.012760158\\
};
\addplot[forget plot, color=mycolor1] table[row sep=crcr] {%
-0.2	0\\
8.2	0\\
};
\addlegendentry{Association Policy 1}

\addplot[ybar, bar width=6, fill=mycolor4, draw=mycolor4, area legend] table[row sep=crcr] {%
1	0.1923193\\
2	0.10632441\\
3	0.067311555\\
4	0.033473644\\
5	0.019894086\\
6	0.016049856\\
7	0.013503527\\
};
\addplot[forget plot, color=mycolor4] table[row sep=crcr] {%
-0.2	0\\
8.2	0\\
};
\addlegendentry{Association Policy 2}

\addplot[ybar, bar width=6, fill=mycolor5, draw=mycolor5, area legend] table[row sep=crcr] {%
1	0.18\\
2	0.1034016\\
3	0.063573055\\
4	0.026109396\\
5	0.013145608\\
6	0.00955824\\
7	0.008029761\\
};
\addplot[forget plot, color=mycolor5] table[row sep=crcr] {%
0.514285714285714	0\\
7.48571428571429	0\\
};
\addlegendentry{Association Policy 3}
\end{axis}
\end{tikzpicture}%
\vspace{0.1cm}
\caption{Beam-pattern error vs training epoch for different \ac{GBS} association policy.}
\label{fig:beamPatternError}
\end{figure}
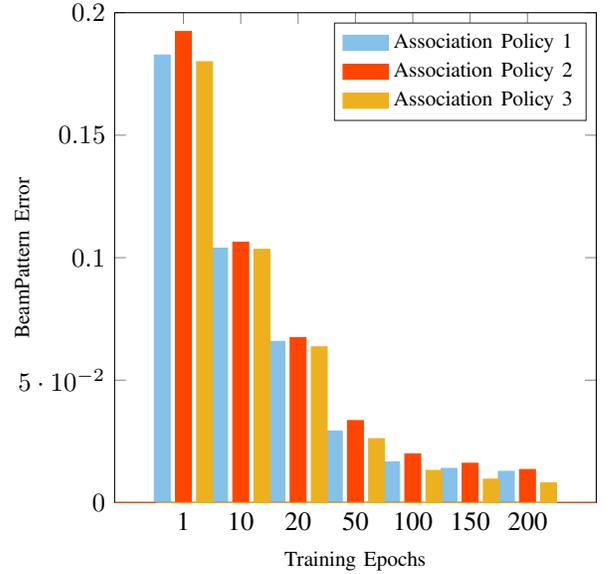
Next, for the communication-only design in \figref{fig:Fig.RadPatt2} and  \figref{fig:FigFar.RadPatt2}, it is observed that the \ac{UAV} antenna main lobe is deployed above the associating \ac{GBS}, and the \ac{UAV}’s transmission power is radiated towards \ac{GBS} to perform the task of communication efficiently. 
Moreover, the beamforming nulls are placed in the neighboring \ac{GBS} directions to minimize the interference at the \ac{GBS} antenna side.

Finally, in \figref{fig:Fig.RadPatt1}, \figref{fig:FigFar.RadPatt1}, it is observed that the \ac{UAV} beamforming direction is deployed between the users and the sensing area.
It is possible to note that when using the ISAC beamforming weights, the $\theta_{\text{3dB}}$ is directed towards the associated \ac{GBS}. In the direction of inferring \ac{GBS} two nulls are placed. On the other hand, when using the sensing weights, the main lobe is directed towards the target while the \ac{GBS} directions are not nulled properly. In this regard, Table \ref{Table:OutputRadPatt} shows the obtained SLL, EIRP and active elements of the desired array pattern at the \ac{UAV} for the two considered points.
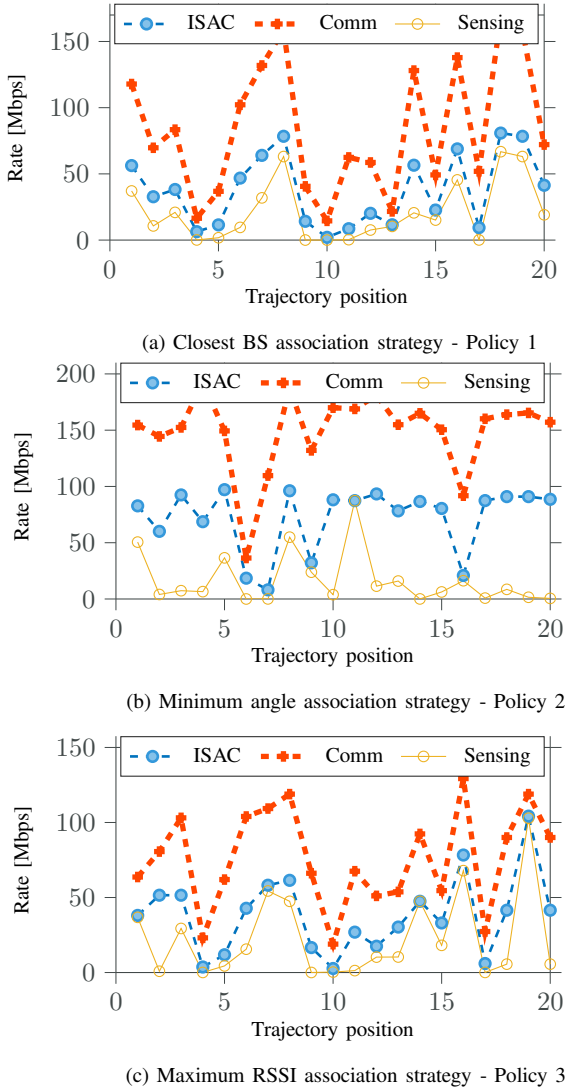
\begin{figure}[t!]
\minipage{0.5\textwidth}
%
%
\definecolor{mycolor1}{rgb}{0.00000,0.44700,0.74100}%
\definecolor{mycolor2}{rgb}{0.52157,0.75686,0.91373}%
\definecolor{mycolor3}{rgb}{0.20392,0.59608,0.85882}%
\definecolor{mycolor4}{rgb}{1.00000,0.27059,0.00000}%
\definecolor{mycolor5}{rgb}{0.92900,0.69400,0.12500}%
\definecolor{mycolor6}{rgb}{0.25882,0.28627,0.28627}%
\begin{tikzpicture}

\begin{axis}[%
width=2.273in,
height=1.179in,
at={(1.006in,0.868in)},
scale only axis,
separate axis lines,
every outer x axis line/.append style={mycolor6},
every x tick label/.append style={font=\color{mycolor6}},
every x tick/.append style={mycolor6},
xmin=0,
xmax=20,
xminorticks=true,
tick align=outside,
xlabel style={font=\color{white!15!black},font=\footnotesize},
xlabel={Trajectory position},
every outer y axis line/.append style={mycolor6},
every y tick label/.append style={font=\color{mycolor6}},
every y tick/.append style={mycolor6},
ymin=0,
ymax=170,
yminorticks=true,
ylabel style={font=\color{white!15!black},font=\footnotesize},
ylabel={Rate [Mbps]},
axis background/.style={fill=white},
legend style={ font=\footnotesize,at={(0.01,0.87)},anchor=south west,fill opacity=0.8, legend columns=-1,legend style={draw=none,column sep=1ex}, draw=white!15!black}
]
\addplot [color=mycolor1, dashed, line width=1.0pt, mark=*, mark options={solid, fill=mycolor2, draw=mycolor3}]
  table[row sep=crcr]{%
1	56.27030673\\
2	32.7331465\\
3	38.20746957\\
4	6.365741607\\
5	11.47860177\\
6	46.81543015\\
7	63.99002247\\
8	78.39514421\\
9	14.19202999\\
10	2.06496402\\
11	8.590434757\\ %
12	20.2078099\\
13	11.5692525\\
14	56.54565516\\
15	22.67062278\\
16	68.87506899\\
17	9.415769564\\
18	80.81159718\\ 
19  78.39514421\\
20  41.40656598\\
};
\addlegendentry{ISAC}

\addplot [color=mycolor4, dashed, line width=2.0pt, mark=+, mark options={solid, mycolor4}]
  table[row sep=crcr]{%
1	117.8637251\\
2	69.72006307\\
3	83.40840615\\
4	16.4966173\\
5	36.83425637\\
6	102.1972648\\
7	131.8816714\\
8	158.6156174\\
9	40.58349275\\
10	14.51691439\\
11	62.43830482\\
12	58.6273315\\
13	21.65106424\\
14	128.0432301\\
15	49.22664656\\
16	137.8357159\\
17	51.926624911\\
18	167.9655756\\
19  158.6156174\\
20  72.12744105\\
};
\addlegendentry{Comm}

\addplot [color=mycolor5, mark=o, mark options={solid, mycolor5}]
  table[row sep=crcr]{%
1	37.09869184\\ 
2	10.69013367\\
3	20.91736756\\
4	0.1104395202\\
5	1.752764543\\
6	9.577598226\\
7	31.88186864\\
8	63.09363458\\
9	 0.02614188635\\
10	0.01043244153\\
11	0.3090629012\\
12	7.600523855\\
13	10.32435048\\
14	20.58256023\\
15	15.11538705\\
16	45.45532559\\
17	0.104844448\\
18	66.7657415\\
19  63.09363458\\
20  19.09443919\\
};
\addlegendentry{Sensing}

\end{axis}
\end{tikzpicture}%



  \subcaption{Closest BS association strategy - Policy 1}\label{fig:}
\endminipage\hfill
\vspace{0.2cm}
\minipage{0.5\textwidth}
%
%
\definecolor{mycolor1}{rgb}{0.00000,0.44700,0.74100}%
\definecolor{mycolor2}{rgb}{0.52157,0.75686,0.91373}%
\definecolor{mycolor3}{rgb}{0.20392,0.59608,0.85882}%
\definecolor{mycolor4}{rgb}{1.00000,0.27059,0.00000}%
\definecolor{mycolor5}{rgb}{0.92900,0.69400,0.12500}%
\definecolor{mycolor6}{rgb}{0.25882,0.28627,0.28627}%
\begin{tikzpicture}

\begin{axis}[%
width=2.273in,
height=1.179in,
at={(1.006in,0.868in)},
scale only axis,
separate axis lines,
every outer x axis line/.append style={mycolor6},
every x tick label/.append style={font=\color{mycolor6}},
every x tick/.append style={mycolor6},
xmin=0,
xmax=20,
xminorticks=true,
tick align=outside,
xlabel style={font=\color{white!15!black},font=\footnotesize},
xlabel={Trajectory position},
every outer y axis line/.append style={mycolor6},
every y tick label/.append style={font=\color{mycolor6}},
every y tick/.append style={mycolor6},
ymin=0,
ymax=200,
yminorticks=true,
ylabel style={font=\color{white!15!black},font=\footnotesize},
ylabel={Rate [Mbps]},
axis background/.style={fill=white},
legend style={ font=\footnotesize,at={(0.01,0.87)},anchor=south west,fill opacity=0.8, legend columns=-1,legend style={draw=none,column sep=1ex}, draw=white!15!black}
]
\addplot [color=mycolor1, dashed, line width=1.0pt, mark=*, mark options={solid, fill=mycolor2, draw=mycolor3}]
  table[row sep=crcr]{%
1	82.75407143\\
2	60.10429917\\
3	92.46658106\\
4	68.68860233\\
5	97.15939094\\
6	18.48776401\\
7	8.13692273\\
8	96.1753268\\ 
9	32.1628078\\
10	88.13605089\\
11	87.38743217\\
12	93.35120952\\
13	78.30020941\\
14	86.63947957\\
15	80.44432345\\
16	20.89113595\\
17	87.43891199\\
18	90.95812241\\
19	90.95812241\\
20	88.56794612\\
};
\addlegendentry{ISAC}

\addplot [color=mycolor4, dashed, line width=2.0pt, mark=+, mark options={solid, mycolor4}]
  table[row sep=crcr]{%
1	154.5375993\\
2	144.4345245\\
3	152.425256\\
4	192.3337083\\
5	149.3549125\\
6	36.40847336\\
7	109.6240172\\
8	187.8629857\\
9	132.0351566\\
10	169.9560124\\
11	168.9047029\\
12	180.9134945\\
13	154.8675515\\
14	164.9873189\\
15	150.42174\\
16	91.76668504\\
17	160.1241243\\
18	163.8329259\\
19	165.3224744\\
20	157.1066238\\
};
\addlegendentry{Comm}

\addplot [color=mycolor5, mark=o, mark options={solid, mycolor5}]
  table[row sep=crcr]{%
1	50.56460409\\
2	3.999637002\\
3	7.379384102\\
4	6.536745966\\
5	36.55840934\\
6	0.009531477\\
7	0.075693851\\
8	54.93704191\\
9	23.8142837\\
10	3.837476755\\
11	87.32334618\\
12	11.41687163\\
13	15.94567365\\
14	0.053591248\\
15	6.252019575\\
16	16.2370156\\
17	0.79306778\\
18	8.569056111\\
19	1.532062612\\
20	0.528322482\\
};
\addlegendentry{Sensing}

\end{axis}
\end{tikzpicture}%
  \subcaption{Minimum angle association strategy - Policy 2}\label{fig:}
\endminipage\hfill
\vspace{0.2cm}
\minipage{0.5\textwidth}%
%
%
\definecolor{mycolor1}{rgb}{0.00000,0.44700,0.74100}%
\definecolor{mycolor2}{rgb}{0.52157,0.75686,0.91373}%
\definecolor{mycolor3}{rgb}{0.20392,0.59608,0.85882}%
\definecolor{mycolor4}{rgb}{1.00000,0.27059,0.00000}%
\definecolor{mycolor5}{rgb}{0.92900,0.69400,0.12500}%
\definecolor{mycolor6}{rgb}{0.25882,0.28627,0.28627}%
\begin{tikzpicture}

\begin{axis}[%
width=2.273in,
height=1.179in,
at={(1.006in,0.868in)},
scale only axis,
separate axis lines,
every outer x axis line/.append style={mycolor6},
every x tick label/.append style={font=\color{mycolor6}},
every x tick/.append style={mycolor6},
xmin=0,
xmax=20,
xminorticks=true,
tick align=outside,
xlabel style={font=\color{white!15!black},font=\footnotesize},
xlabel={Trajectory position},
every outer y axis line/.append style={mycolor6},
every y tick label/.append style={font=\color{mycolor6}},
every y tick/.append style={mycolor6},
ymin=0,
ymax=150,
yminorticks=true,
ylabel style={font=\color{white!15!black},font=\footnotesize},
ylabel={Rate [Mbps]},
axis background/.style={fill=white},
legend style={ font=\footnotesize,at={(0.01,0.87)},anchor=south west,fill opacity=0.8, legend columns=-1,legend style={draw=none,column sep=1ex}, draw=white!15!black}
]
\addplot [color=mycolor1, dashed, line width=1.0pt, mark=*, mark options={solid, fill=mycolor2, draw=mycolor3}]
  table[row sep=crcr]{%
1	38.0599486\\
2   51.5250185\\
3   51.52282078\\
4   3.693434867\\
5   11.84573475\\
6   42.8144369\\
7  58.18276379\\
8  61.44884568\\
9  16.63994307\\
10  2.696502824\\
11 26.91736053\\
12  17.54500948\\
13  30.30871671\\
14  47.48892532\\
15  33.04591688\\
16  78.30471254\\
17  6.003860672\\
18  41.52524943\\
19  104.2304144\\
20  41.52524943\\
};
\addlegendentry{ISAC}

\addplot [color=mycolor4, dashed, line width=2.0pt, mark=+, mark options={solid, mycolor4}]
  table[row sep=crcr]{%
1	63.65669114\\
2 80.60506637\\
3 103.0928449\\
4 23.11240273\\
5  61.96171875\\
6  103.9969298\\
7  109.2772133\\
8	118.9008904\\
9	66.17567917\\
10	18.93492299\\
11	67.53110706\\
12  51.0813217\\
13  53.76409916\\
14 92.38156228\\
15  54.59287044\\
16  129.2562521\\
17  27.34479186\\
18  90.00276501\\
19  118.9008904\\
20  90.00276501\\
};
\addlegendentry{Comm}

\addplot [color=mycolor5, mark=o, mark options={solid, mycolor5}]
  table[row sep=crcr]{%
1	36.77206617\\
2	0.8606703532\\
3	29.36950659\\
4	0.05523822638\\
5	4.41344585\\
6	15.55977395\\
7	54.36518763\\
8	47.46055758\\
9	0.04055224541\\
10	0.3029531607\\
11	1.29041296\\
12	10.284321\\
13	10.3491101\\
14	47.46055758\\
15	18.11608986\\
16	67.9033631\\
17	0.07633267664\\
18	5.594706767\\
19 102.8278484\\
20 5.594706767\\
};
\addlegendentry{Sensing}

\end{axis}
\end{tikzpicture}%



  \subcaption{Maximum RSSI association strategy - Policy 3}\label{fig:}
\endminipage
\caption{Instantaneous rate under different requirements: sensing, communication, and ISAC} 
        \label{fig:Rate_vs_trajectory}
\end{figure}
\figref{fig:AzimuthElevationTransmitBeamPattern}-left shows the transmit beampattern obtained by the DNN beamformer. The dashed line specifies the ISAC azimuth profile obtained by summing the communication and the sensing profile. Its counterpart, the transmit beampattern obtained by the beamforming optimizer, is shown in 
\figref{fig:AzimuthElevationTransmitBeamPattern}-right. The dashed line specifies the ISAC azimuth profile obtained by summing the communication and the sensing profile.
It is observed that for the communication case, the pattern is formed to meet the SINR requirements with minimum power, while, for the sensing case, the pattern is created such that the resultant matching error can be minimized.

The proposed \ac{DNN} solution can predict the beamforming weights with a significantly reduced time compared to running the beampattern optimizer for any point of the possible \ac{UAV} trajectories. The training length and time for the proposed model are reported in Table \ref{table:TrainingTime}.

\figref{fig:beamPatternError} shows the beampattern error along the \ac{DNN} training epochs. The beampattern error diminishes with increasing training epochs, showing a convergent learning process for all three association policies. It means a stable beampattern at the \ac{UAV} antenna is learned.

\subsection{GBS association policies evaluation}
To assess the \ac{GBS} association policy performance, we consider the approaches described in Sec.~\ref{BS_Association}. For the sake of clarity, we adopt the following notation:
\begin{itemize}
    \item Policy 1 refers to choosing the closest \ac{GBS};
    \item Policy 2 refers to choosing the \ac{GBS} with azimuth angle closest to the target;
    \item Policy 3 refers to choosing the \ac{GBS} experiencing the highest \ac{SINR}.
\end{itemize}
\begin{figure}[tp]
\centering
\input{Figures/EirpDistribution_vs_policies.tex}
\vspace{0.1cm}
\caption{\Ac{ECDF}, indicated with $\mathsf{ECDF}(c)$, for the EIRP under different GBS association strategies.}\label{fig:EIRPassociationPolicies}
\end{figure}

\begin{figure}[tp]
\centering
\includegraphics[width=\linewidth]{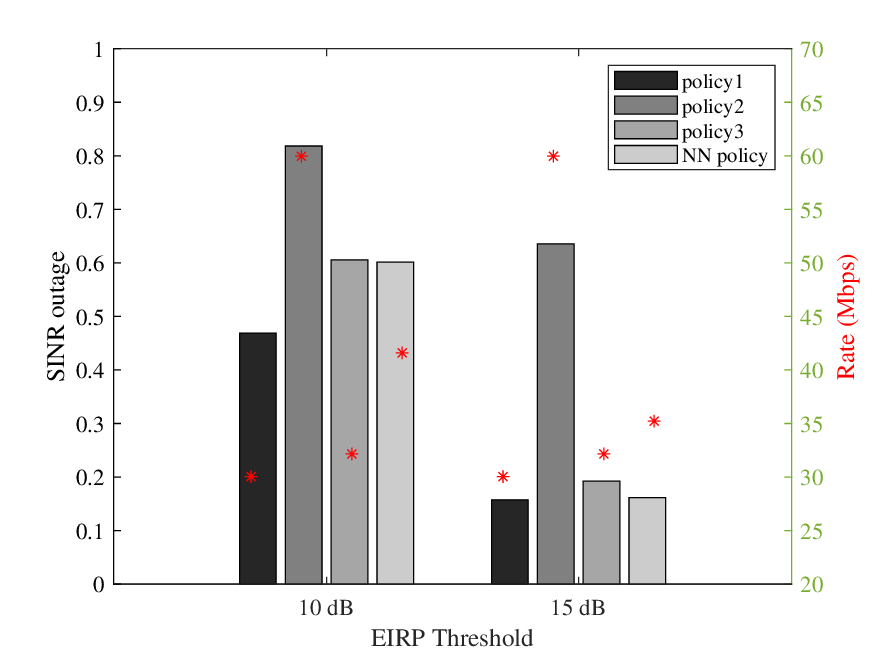}
\vspace{0.1cm}
\caption{\Ac{EIRP} outage (gray bars) and communication average rate (red markers) for \ac{EIRP} beampattern thresholds of $10\,\mathrm{dBm}$ and $15\,\mathrm{dBm}$.}\label{fig:EIRPvsRate}
\end{figure}
It is also interesting to discuss the instantaneous communication rates achieved by different users in \figref{fig:Rate_vs_trajectory}. Results in \figref{fig:Rate_vs_trajectory} are coherent with what was found in  \figref{fig:Radpatt_close} and \figref{fig:Radpatt_far}. The beampattern optimized for communication provides the highest instantaneous rate along the \ac{UAV} trajectory, while the ISAC design can balance the tradeoff between communication and sensing performances.

However, evaluating the best \ac{GBS} association policy includes the evaluation of the trade-off between constraints \ref{eq:cond3_Opt_problem1_discrete} and \ref{eq:cond5_Opt_problem1_discrete} in the optimization problem.
Condition \ref{eq:cond3_Opt_problem1_discrete} imposes to satisfy a maximum transmission power/\ac{EIRP} at the antenna side to limit the energy consumed by the \ac{UAV}.
The \ac{EIRP}, in turn, has to be high enough to satisfy the \ac{SINR} condition \ref{eq:cond5_Opt_problem1_discrete} at the \ac{GBS} side to be able to decode the \ac{UAV} signal.
\figref{fig:EIRPassociationPolicies} shows the \ac{EIRP} distribution during the \ac{UAV} paths under the three different association strategies.
While policy 2 helps to reach higher values of communication rate, policies 1 and 3 have lower \ac{EIRP} values, leading to less power dissipation at the \ac{UAV} antenna side. 

\figref{fig:EIRPvsRate} shows the average \ac{EIRP} versus the achievable rate for the \ac{UAV}, comparing the three benchmark association policies defined and the proposed NN \ac{GBS} association. The comparison is performed considering two different EIRP thresholds at the transmitter, $10$ and $15\, \mathrm{dBm}$, respectively. We can see that the DNN-based association strategy led to a better rate performance with a lower EIRP outage than all other association strategies, which shows the great potential of our proposed approach in the \ac{ISAC} \ac{UAV}-enabled communication scenario.

\section{Conclusions}
\label{sec:conclusions}
In this paper, we proposed \ac{DNN} solution for the  joint \ac{ISAC} beamforming and \ac{UAV}-\ac{GBS} association problem.
The derived optimization problem accounts for \ac{UAV} trajectory and orientation, sensing and communication beamforming weights, and \ac{GBS} selection.
Thus, the addressed optimization problem is comprehensive, considering various factors such as the trajectory and orientation of \acp{UAV}, beamforming weights for sensing and communication, and selecting suitable \acp{GBS}. Given the intricacies involved, particularly the nonconvex nature of the cost function and constraints, our approach leverages the power of two distinct DNNs. Our results demonstrate the considerable potential of our proposed approach in the context of \ac{UAV}-enabled communication scenarios. Firstly, our solution showcases the capability to predict beamforming weights with significantly reduced computational time compared to state-of-the-art beampattern optimizers. Secondly, by utilizing a second DNN to determine the most suitable GBS association at each point along the \ac{UAV} trajectory, our approach proves superior to traditional GBS association rules in terms of \ac{EIRP}, \ac{SINR} performance and computational speed.

\section*{Acknowledgment}
{\footnotesize The research was partially funded by the Luxembourg National Research Fund (FNR) under the project SmartSpace (C21/IS/16193290).}

\begin{acronym} 
\acro{3G}{Third Generation}
\acro{4G}{Fourth Generation}
\acro{5G}{Fifth Generation}
\acro{6G}{sixth generation}
\acro{3GPP}{3rd Generation Partnership Project}
\acro{BB}{Base Band}
\acro{BBU}{Base Band Unit}
\acro{BER}{Bit Error Rate}
\acro{BH}{Backhaul}
\acro{BPP}{Binomial Point Process}
\acro{BS}{Base Station}
\acro{BW}{bandwidth}
\acro{C-RAN}{Cloud Radio Access Networks}
\acro{CAPEX}{Capital Expenditure}
\acro{CBR}{case-based reasoning}
\acro{ECDF}{empirical cumulative distribution function}
\acro{CRLB}{Cramer-Rao lower bound}
\acro{CoMP}{Coordinated Multipoint}
\acro{CMDP}{Constrained Markov Decision Process}
\acro{CPRI}{common public radio interface }
\acro{CU}{Centralized Unit}
\acro{D2D}{device-to-Device}
\acro{DAC}{digital-to-Analog Converter}
\acro{DAS}{distributed Antenna Systems}
\acro{DBA}{dynamic Bandwidth Allocation}
\acro{DFRC}{dual function radar communication}
\acro{DL}{Downlink}
\acro{DOF}{degrees of freedom}
\acro{DNN}{deep neural network}
\acro{DQN}{deep Q-network}
\acro{DDQN}{Double Deep Q-Network}
\acro{DQfD}{Deep Q-Learning from Demonstration}
\acro{DRL}{deep reinforcement learning}
\acro{DU}{Distributed Unit}
\acro{EIRP}{effective isotropic radiated power}
\acro{FBMC}{Filterbank Multicarrier}
\acro{FEC}{Forward Error Correction}
\acro{FFNN}{feed forward neural network}
\acro{FH}{Fronthaul}
\acro{FL}{Federated Learning}
\acro{FFR}{Fractional Frequency Reuse}
\acro{FSO}{Free Space Optics}
\acro{GBS}{ground base station}
\acro{GSM}{Global System for Mobile Communications}
\acro{HAP}{High Altitude Platform}
\acro{HetNet}{Heterogeneous Network}
\acro{HL}{Higher Layer}
\acro{HARQ}{Hybrid-Automatic Repeat Request}
\acro{ISAC}{Integrated Sensing and Communication}
\acro{KL}{Kullback-Leibler}
\acro{KPI}{Key Performance Indicator}
\acro{kg}{Kilogramm}
\acro{HPBW}{Half-Power-Beam-Width}
\acro{IoT}{Internet of Things}
\acro{LAN}{Local Area Network}
\acro{LAP}{Low Altitude Platform}
\acro{LfD}{Learning From Demonstration}
\acro{LL}{Lower Layer}
\acro{LoS}{line-of-sight}
\acro{LTE}{Long Term Evolution}
\acro{LTE-A}{Long Term Evolution Advanced}
\acro{LP}{Linear Programming}
\acro{MAC}{Medium Access Control}
\acro{MAP}{Medium Altitude Platform}
\acro{MC}{Monte Carlo}
\acro{MDP}{Markov Decision Process}
\acro{ML}{Medium Layer}
\acro{MME}{Mobility Management Entity}
\acro{mmWave}{millimeter Wave}
\acro{MIMO}{Multiple Input Multiple Output}
\acro{ML}{Machine Learning}
\acro{MSE}{Mean Square Error}
\acro{NFP}{Network Flying Platform}
\acro{NFPs}{Network Flying Platforms}
\acro{NN}{Neural Network}
\acro{NLoS}{Non-Line of Sight}
\acro{RL}{Reinforcement Learning}
\acro{NR}{New Radio}
\acro{OFDM}{Orthogonal Frequency Division Multiplexing}
\acro{PAM}{Pulse Amplitude Modulation}
\acro{PAPR}{Peak-to-Average Power Ratio}
\acro{PDF}{Probability Density Function}
\acro{PER}{Prioritized Experience Replay}
\acro{PGW}{Packet Gateway}
\acro{PHY}{physical layer}
\acro{PP}{Poisson Process}
\acro{PSO}{Particle Swarm Optimization}
\acro{PTP}{Poin to Point}
\acro{QAM}{Quadrature Amplitude Modulation}
\acro{QoE}{Quality of Experience}
\acro{QoS}{Quality of Service}
\acro{QPSK}{Quadrature Phase Shift Keying}
\acro{RF}{Radio Frequency}
\acro{RIS}{Reconfigurable Intelligent Surface}
\acro{RN}{Remote Node}
\acro{RAU}{Remote Access Unit}
\acro{RAN}{Radio Access Network}
\acro{RRH}{Remote Radio Head}
\acro{RRU}{Remote Radio Unit}
\acro{RRC}{Radio Resource Control}
\acro{RRU}{Remote Radio Unit}
\acro{RSS}{Received Signal Strength}
\acro{RSRP}{Reference Signals Received Power}
\acro{RU}{Remote Unit}
\acro{SCBS}{Small Cell Base Station}
\acro{SDN}{Software Defined Network}
\acro{SINR}{Signal-to-Noise-plus-Interference Ratio}
\acro{SIR}{Signal-to-Interference Ratio}
\acro{SLL}{side lobe level}
\acro{SNR}{signal-to-noise ratio}
\acro{SON}{Self-organising Network}
\acro{TETRA}{Trans-European Trunked Radio}
\acro{THz}{Terahertz}
\acro{TL}{Transfer Learning}
\acro{TD}{Temporal Difference}
\acro{TDD}{Time Division Duplex}
\acro{TD-LTE}{Time Division LTE}
\acro{TDM}{Time Division Multiplexing}
\acro{TOA}{time of arrival}
\acro{TDMA}{Time Division Multiple Access}
\acro{UE}{User Equipment}
\acro{UAV}{unmanned aerial vehicle}
\acro{ULA}{uniform linear array}
\acro{UPA}{uniform planar array}
\acro{UPA}{Uniform Planar Square Array}

\acro{SAR}{synthetic-aperture radar}
\end{acronym}

\bibliographystyle{IEEEtran}
\bibliography{References.bib}

\end{document}